\documentclass{aa}  

\usepackage{natbib,twoopt}
\bibpunct{(}{)}{;}{a}{}{,}             %% natbib format for A&A and ApJ

\usepackage[table]{xcolor}
\usepackage{color, colortbl}
\definecolor{darkpastelgreen}{rgb}{0.01,0.75,0.24}
\definecolor{lightgreen}{rgb}{0.56, 0.93, 0.56}
\definecolor{brightgreen}{rgb}{0.4, 1.0, 0.0}
\definecolor{carrotorange}{rgb}{0.93, 0.57, 0.13}
\definecolor{darkorange}{rgb}{1.0, 0.55, 0.0}
\definecolor{bostonuniversityred}{rgb}{0.8, 0.0, 0.0}
\definecolor{cadmiumred}{rgb}{0.89, 0.0, 0.13}
\definecolor{candyapplered}{rgb}{1.0, 0.03, 0.0}
\usepackage{hhline}
\usepackage{array,multirow,graphicx}
\usepackage{lscape}

\usepackage{tablefootnote}
\usepackage[flushleft]{threeparttable}

\usepackage{txfonts}
\usepackage{hyperref}
\begin{document}

   \title{ATMOSPHERIX: III- Estimating the C/O ratio and molecular dynamics at the limbs of WASP-76 b with SPIRou 
             }

   \author{Thea Hood \inst{1} \thanks{Based on observations obtained at the Canada-France-Hawaii Telescope (CFHT) which is operated from the summit of Maunakea by the National Research Council of Canada, the Institut National des Sciences de l'Univers of the Centre National de la Recherche Scientifique of France, and the University of Hawaii. The observations at the Canada-France-Hawaii Telescope were performed with care and respect from the summit of Maunakea which is a significant cultural and historic site. }
          \and Florian Debras \inst{1}
          \and Claire Moutou \inst{1}
          \and Baptiste Klein \inst{2}
          \and Pascal Tremblin \inst{3}
          \and Vivien Parmentier \inst{4}
          \and Andres Carmona \inst{5}
          \and Annabella Meech \inst{2}
          \and Olivia Vénot \inst{6}
          \and Adrien Masson \inst{7}
          \and Pascal Petit \inst{1}
          \and Sandrine Vinatier \inst{7}
          \and Eder Martioli \inst{8,9}
          \and Flavien Kiefer \inst{8}
          \and Martin Turbet \inst{10,11}
          \and the ATMOSPHERIX consortium
          }

   \institute{IRAP, UMR5277 CNRS, Université de Toulouse, UPS, Toulouse, France\\
              \email{thea.hood@irap.omp.eu}
         \and Department of Physics, University of Oxford, OX13RH, Oxford, UK\
         \and Universite Paris-Saclay, UVSQ, CNRS, CEA, Maison de la Simulation, 91191, Gif-sur-Yvette, France\
         \and Université Côte d'Azur, Observatoire de la Côte d'Azur, CNRS, Lagrange, CS 34229, Nice, France\
         \and Université Grenoble Alpes, CNRS, IPAG, 38000 Grenoble, France\
         \and Université de Paris Cité and Univ Paris Est Creteil, CNRS, LISA, F-75013 Paris, France\
         \and LESIA, Observatoire de Paris, Université PSL, Sorbonne Université, Université Paris Cité, CNRS, 5 place Jules Janssen, 92195 Meudon, France \
         \and Institut d'Astrophysique de Paris, CNRS, UMR 7095, Sorbonne
Universit\'{e}, 98 bis bd Arago, 75014 Paris, France\
         \and Laborat\'{o}rio Nacional de Astrof\'{i}sica, Rua Estados
Unidos 154, 37504-364, Itajub\'{a} - MG, Brazil \
         \and Laboratoire de Météorologie Dynamique/IPSL, CNRS, Sorbonne Université, Ecole Normale Supérieure, Université PSL, Ecole Polytechnique, Institut Polytechnique de Paris, 75005 Paris, France
         \and Laboratoire d’astrophysique de Bordeaux, Univ. Bordeaux, CNRS, B18N, allée Geoffroy Saint-Hilaire, 33615 Pessac, France
             }

   \date{Received date /
   Accepted date }

% \abstract{}{}{}{}{} 
% 5 {} token are mandatory
 
  \abstract{ Measuring the abundances of C- and O-bearing species in exoplanet atmospheres enables us to constrain the C/O ratio, that contains indications about the planet formation history. With a wavelength coverage going from 0.95 to 2.5 microns, the high-resolution (R~$\sim$~70~000) spectropolarimeter SPIRou can detect spectral lines of major bearers of C and O in exoplanets. Here we present our study of SPIRou transmission spectra of WASP-76 b acquired for the ATMOSPHERIX program. We applied the publicly available data analysis pipeline developed within the ATMOSPHERIX consortium, analysing the data using 1-D models created with the \texttt{\texttt{petitRADTRANS}} code, with and without a grey cloud deck. We report the detection of H$_2$O and CO at a Doppler shift of around -6~km.s$^{-1}$, consistent with previous observations of the planet. Finding a deep cloud deck to be favoured, we measured in mass mixing ratio (MMR) log(H$_2$O)$_{\rm MMR}$ = -4.52 $\pm$ 0.77 and log(CO)$_{\rm MMR}$~=~-3.09 $\pm$ 1.05 consistent with a sub-solar metallicity to more than 1$\sigma$. We report 3$\sigma$ upper limits for the abundances of C$_2$H$_2$, HCN and OH. We estimated a C/O ratio of 0.94 $\pm$ 0.39 ($\sim$ 1.7 $\pm$ 0.7 $\times$ solar, with errors indicated corresponding to the 2$\sigma$ values) for the limbs of WASP-76~b at the pressures probed by SPIRou. We used 1-D \texttt{\texttt{ATMO}} forward models to verify the validity of our estimation. Comparing them to our abundance estimations of H$_2$O and CO, as well as our upper limits for C$_2$H$_2$, HCN and OH, we found that our results were consistent with a C/O ratio between 1 and 2 $\times$ solar, and hence with our C/O estimation. Finally, we found indications of asymmetry for both H$_2$O and CO when investigating the dynamics of their signatures, pointing to a complex scenario involving possibly both a temperature difference between limbs and clouds being behind the asymmetry this planet is best known for.} 

   \keywords{Planets and satellites: atmospheres -- Planets and satellites: individual (WASP-76 b) -- Techniques: spectroscopic -- Methods: data analysis
               }

    \titlerunning{ATMOSPHERIX: III- Estimating the C/O ratio and molecular dynamics at the limbs of WASP-76 b with SPIRou}
    \authorrunning{T. Hood et al.}

   \maketitle
%
%-------------------------------------------------------------------

\section{Introduction}

With thousands of exoplanets now discovered, a major next step in their study is the characterisation of their atmospheres. By doing so, we can obtain information about their chemical and physical properties. This has already been performed for over a hundred planets\footnote{According to \url{http://research.iac.es/proyecto/exoatmospheres/index.php}, a database referencing exoplanet atmospheres studies} using data acquired using space-based and/or ground-based telescopes (see notably a review in \citealt{Guillot2010}, as well  as the first results from James Webb:  \citealt{Taylor2023,JWST2023}). Such information can help to constrain the scenarios in which the exoplanets formed and evolved in time (e.g., \citealt{Madhu2017}).

Transmission spectroscopy is an important tool in the characterisation of exoplanetary atmospheres. The variations in stellar light due to absorption by the atmosphere of an exoplanet as it passes in front of its host star result in spectra that contain information on both the temperature and chemical properties of the atmosphere at the limbs of the exoplanet (commonly referred to as morning and evening). From the measured properties as well as any differences between the morning and evening limbs, properties concerning the 3-D structure of the atmosphere can be inferred \citep{Pluriel2023}. From the ground, the Doppler shift of a planet observed with high-resolution spectroscopy also enables the inference of dynamical processes in the atmosphere (e.g., \citealt{Snellen2010,Flowers2019}).

One of the most recent near-infrared (nIR) ground-based high-resolution instruments is SPIRou (Spectro-Polarimètre InfraRouge; \citealt{Donati2020}), a fibre-fed cryogenic échelle spectrograph. Installed at the Canada-France-Hawaii Telescope in 2018, it has been observing the sky since 2019. SPIRou observes in the near-infrared, with a continuous coverage from 0.95 to 2.5 $\mu$m provided by 49 overlapping diffraction orders. With a resolving power of R $\sim$ 70 000 and a sampling precision of $\sim 2.27$ km.s$^{-1}$ per pixel, SPIRou is one of the best instruments to study volatile species in exoplanetary atmospheres. It has already been successfully used to do so for $\tau$ Boo b \citep{Pelletier2021}, HD 189733 b \citep{Boucher2021,Klein2024} and WASP-127~b \citep{Boucher2023}.\

\begin{table*}
    \centering
    \caption{Adopted WASP-76 System Parameters}
    \label{tab:sysparams}
    {\renewcommand{\arraystretch}{1.42}%
    \begin{tabular}{lccr} % four columns, alignment for each
        \hline
        Stellar parameters & Value & Reference\\
        \hline
        Mass (M$_\odot$) & 1.458$\pm$0.021 & \cite{Ehren2020} \\
        Radius (R$_\odot$) & 1.756$\pm$0.071 & \cite{Ehren2020} \\
        Effective temperature (K) & 6 329$\pm$65 & \cite{Ehren2020} \\
        Metallicity [Fe/H] & 0.366$\pm$0.053 & \cite{Ehren2020} \\
        RV semi-amplitude (m.s$^{-1}$) & 116.02$^{+1.29}_{-1.35}$ & \cite{Ehren2020} \\
        Systemic velocity (V$_{\rm c}$; km.s$^{-1}$) & -1.11$\pm$0.50 & \cite{Soubiran2018} \\
        Limb darkening (Quadratic) & [0.393,0.219] & \cite{Ehren2020} \\
        \hline
        Planetary parameters & Value & Reference\\
        \hline
        Epoch of transit (T$_0$) & 2457273.4191$\pm$0.0005 & \cite{Kokori2022} \\
        Orbital Period (days) & 1.8098806$\pm$0.0000007 & \cite{Kokori2022} \\
        Mass (M$_J$) & 0.894$^{+0.014}_{-0.013}$ & \cite{Ehren2020} \\
        Radius (R$_J$) & 1.856$^{+0.077}_{-0.076}$ & \cite{Ehren2020} \\
        g (m.s$^{-2}$) & 6.4$\pm$0.5 & \cite{Ehren2020} \\
        Planet RV semi-amplitude (km.s$^{-1}$) & 196.52$\pm$0.94 & \cite{Ehren2020} \\
        Semi-major axis (au) & 0.033$\pm$0.0002 & \cite{Ehren2020} \\
        Inclination (deg) & 89.623$^{+0.005}_{-0.034}$ & \cite{Ehren2020} \\
        Eccentricity & 0.0 (fixed) & \cite{Ehren2020} \\
        Argument of the periapsis & 0.0 (fixed) & \cite{Ehren2020} \\
        Transit duration (h) & 3.694 $\pm$ 0.019 & \citet{West2016} \\
        \hline
    \end{tabular}}
\end{table*}

To best exploit the SPIRou observations of exoplanetary atmospheres, the ATMOSPHERIX program (PI: Florian Debras) was created. With a consortium made up of a large French community of specialists in exoplanet atmospheric and stellar observations and simulations, its main goal is to use high-resolution observations to obtain knowledge on exoplanet atmospheric properties. To do so, a pipeline has been developed within the consortium to analyse high-resolution spectroscopic data, optimised for SPIRou data \citep{Klein2024,Debras2024}. Within the program, data of 14 exoplanets has so far been acquired, including data of WASP-76~b.

WASP-76~b \citep{West2016} is an ultra-hot Jupiter (UHJ), with a reported equilibrium temperature of approximately 2200K. Having been observed with both space- and ground-based instruments, a variety of atomic and molecular detections have been reported for the atmosphere of this planet. Data acquired with HST has led to detections of H$_2$O and Na, and tentatively TiO and VO \citep{Tsiaras2018,Fisher2018,vonEssen2020,Edwards2020,Fu2021}. Furthermore, Spitzer data showed a strong CO emission feature \citep{Fu2021}. Sodium has also been detected using data from HARPS \citep{Seidel2019,Zak2019}, ESPRESSO \citep{Tabernero2021,Kesseli2022,AzevedoSilva2022}, Subaru/HDS \citep{Kawauchi2022}, MAROON-X \citep{Pelletier2023} and GRACES \citep{Deibert2023}. The ESPRESSO data was also used to detect Li, Mg, Ca$^+$, Mn, K, Fe, V, Cr, Ni, Sr$^+$, Ba$^+$ and tentatively H, K and Co \citep{Ehren2020,Tabernero2021,Kesseli2022,AzevedoSilva2022,Gandhi2022}. HARPS data also led to the detection of Fe \citep{Kesseli2021}. Detections of Ca$^+$ and Fe were also made using GRACES, as well as tentative detections of Li, K, Cr and V \citep{Deibert2021,Deibert2023}. MAROON-X data also led to detections of Fe, Ca$^+$, Cr, Li, H, V, VO, Mn, Ni, Mg, Ca, K, Ba$^+$, tentatively detect O and Fe$^+$, and find possible evidence of cold trapping of materials in the night side of the planet \citep{Pelletier2023}. CARMENES transmission data has led to reported detections of Ca$^+$, OH, HCN, H$_2$O and tentatively NH$_3$ \citep{CasasayasBarris2021,Landman2021,SanchezLopez2022}. Emission data obtained with CRIRES+ has led to a detection of CO \citep{Yan2023}. We have summarised the different detections for WASP-76~b in Table \ref{tab:chemlist_table}. There are currently no published papers retrieving abundance values for both water and CO for the atmosphere of WASP-76~b, though a similar work to that presented has been started using IGRINS data, with the first results having been presented at Exoplanets IV (see poster presentation of \citealt{Mansfield2022}). The iron detection by \cite{Ehren2020} showed an asymmetry between the morning and evening terminators of WASP-76~b. The detection signal seemed to indicate iron present on the evening limb of the planet but absent on the morning side. This asymmetric iron signal was confirmed by \cite{Kesseli2021} using archival HARPS data. Different explanations were given for this phenomenon: \cite{Ehren2020} put forward a condensation of iron on the night-side of the planet leading to a lack of iron on the morning side, while \cite{Wardenier2021} and \cite{Savel2022} showed that it could also be explained by respectively a substantial temperature asymmetry between limbs and the presence of high-altitude optically thick clouds on the morning terminator.

In this paper, we present our analysis of SPIRou acquired data of WASP-76~b. We will first describe the WASP-76~b data acquired by SPIRou and the reduction process used in preparation for its analysis in section \ref{sec:Obs}. We then describe the models and methods with which these reduced data were analysed in section \ref{sec:Methods}. The results are presented in section \ref{sec:Results}, including notably molecular detections and dynamical effects. We finally discuss these results in section \ref{sec:Discussion}, giving an estimate of the C/O ratio of the planet. Section \ref{sec:concl} is devoted to the conclusion. 

%--------------------------------------------------------------------
\section{Observations and data reduction}
\label{sec:Obs}
\subsection{Observations}
\begin{figure}
    \includegraphics[width=\linewidth]{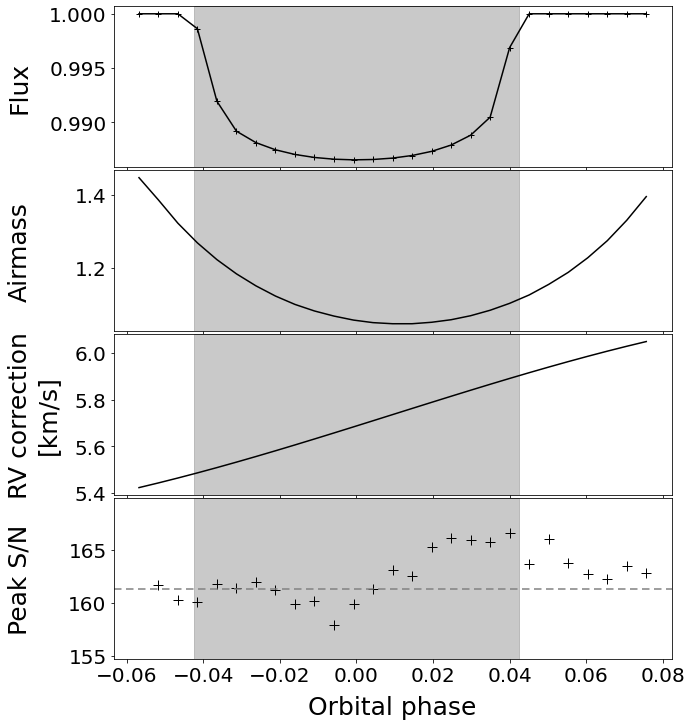}
    \caption{In order from top to bottom are represented the continuum-normalised light-curve, the airmass, the radial velocity (RV) correction (containing RV contributions of the barycentric Earth motion and the stellar systemic velocity), and peak SNR per velocity bin over the course of the transit of WASP-76~b as a function of the orbital phase, computed using the parameters listed in Table \ref{tab:sysparams}. The grey band indicates the transit duration of the planet, from first to fourth contact. The horizontal dashed line in the bottom panel represents the average value of the peak SNR.}
    \label{fig:propsofWASP-76}
\end{figure}

The UHJ WASP-76 b was observed as part of the ATMOSPHERIX program during the night of October 31$^{\rm st}$ in 2020. Lasting approximately 6 hours, the observation consisted of 28 exposures and captured a full transit of the planet, with the first four and last seven exposures being out-of-transit and the rest in between being in-transit. The 2$^{\rm nd}$ spectrum was removed due to this second exposure having been aborted, having an integration time of only 33.432 s (with 774.497 s being the requested integration time). The remaining 27 spectra were used for our study. The adopted parameters of the observed star-planet system are listed in Table \ref{tab:sysparams}. Over the whole observation, the peak signal-to-noise ratio (SNR) per 2.3 km.s$^{-1}$ velocity bin varies between 127 to 167 (with a mean of $\sim$ 161), and the airmass between 1.05 and 1.45. The transit curve as well as the variations in airmass, the radial correction values and registered peak signal-to-noise ratio values over the different exposures can be seen in Figure \ref{fig:propsofWASP-76}. 

\subsection{Data reduction} \label{DataRed}
The data obtained of WASP-76~b were reduced using the version 0.6.132 of \texttt{\texttt{APERO}} (A PipelinE to Reduce Observations), SPIRou’s data reduction software \cite[DRS;][]{Cook2022}. 
% Among others, it provides a telluric corrected format of the data, a correction that performed using a PCA and a library of telluric standards following the method described in \cite{Artigau2014}. It is this format of the data from WASP-76~b that was used as a starting point for studying this planet with SPIRou data. 
\texttt{\texttt{APERO}} applies the optimal extraction method of \citet{Horne1986} to extract each individual exposure from the H4RG detector \citep{Artigau2018}. The wavelength solution is obtained by combining calibration exposures of a UNe hollow-cathode lamp and a thermally-stabilised Fabry-Pérot etalon \citep{bauer2015,hobson2021}. \texttt{\texttt{APERO}} performs a correction of the telluric contamination using a method, summarised in \citet[][see Section~8]{Cook2022}, which will be presented in a forthcoming paper (Artigau et al. in prep.). This technique applies \texttt{TAPAS} \citep{bertaux2014} to preclean telluric absorption and the low level residuals are removed in using a data set of spectra of hot stars observed in different atmospheric conditions to build a residual  models as a function of few parameters (optical depths of H$_2$O and of dry components). Note that the deepest telluric lines (relative absorption larger than 90\%) are masked out by the pipeline as the low amount of transmitted flux will most likely result in an inaccurate telluric correction. Our input sequence of spectra contains the blaze- and telluric-corrected spectra.  As the quality of this telluric correction improves with the number of epochs used to create the correction template and there have been fairly few observations of WASP-76~b, there is a high probability of telluric contamination remaining in the spectra.

We then apply the ATMOSPHERIX pipeline\footnote{ \url{https://github.com/baptklein/ATMOSPHERIX_DATA_RED}} to reduce transmission spectroscopy data \citep{Klein2024}.
% , was applied to this data with the aim of maximising the atmospheric signal in the measured spectra. 
It is made up of different steps, summarised as follows.
\begin{enumerate}
    
    \item The spectra are all aligned in the stellar rest frame, from which is created a master-out spectrum I$_\mathrm{ref}$ by averaging out-of-transit exposures. This master-out spectrum is then moved back to the geocentric frame and linearly matched in flux to each of the observed spectra, which are each divided by its best-fitting solution. The resulting spectra are then divided by a second master-out spectrum, created by averaging the out-of-transit parts of these spectra, now in the Earth rest frame, to provide an additional correction of the tellurics. The option of correcting for planet-induced distortions of the stellar line profile \citep{Chiavassa&Brogi2019} was not used, as the star has a low rotational velocity.
    
    \item A normalisation of each of the resulting spectra is then performed using an estimate of the noise-free continuum, calculated using a rolling mean window. This is followed by a 5-$\sigma$ clipping being applied to remove outliers. This two-step process is then repeated until there are no more outliers flagged in the data.

    \item As some pixels with high temporal variance might remain, we calculated the variance for each pixel and applied an iterative parabolic fit to the pixel variance distribution. We considered pixels further than 5-$\sigma$ from the fit as outliers, and masked them out for the rest of the data reduction process.

    \item The next step is optional, consisting of detrending with the geometric airmass in log-space accordingly to \cite{Brogi2018}. We chose to use it, performing a quadratic detrending of the resulting spectra, as it provides an additional correction for residual tellurics.

    \item Finally, we further correct remaining correlated noise through the use of a principal component analysis (PCA). The number of principal components (PC) to remove per order is automatically chosen by comparing the PC eigenvalues to the variance of a map of white noise with similar dispersion as our data. 
    PCs with eigenvalues larger than the white noise mean eigenvalue are discarded. For our WASP-76~b data, between 1 and 3 PCs were removed for each order. 

\end{enumerate}

Unfortunately, applying a PCA to the data also affects the planetary signal, degrading it. Previous works have tried to take into account at best this effect so as to optimise the analysis of the data \citep{BrogiLine2019,Gibson2020,Boucher2021,Pelletier2021}. To take into account the degradation of our data of WASP-76~b, we implemented the method of \cite{Gibson2022}, being the fastest method for PCA. For this, during the data reduction process, we keep for each order a matrix U of the removed eigenvectors, assumed to be associated to correlated noise. These are subsequently used to prepare the synthetic spectra for analysing the data to degrade them coherently with the real signal (see section \ref{ModelPrep}).

%%%%%%%%%%%%%%%%%%%%%%%%%%%%%%%%%%%%%%%%%%%%%%%%%%%%%%%%%%%%%%%%%%%%%%%%%%%%%%%
%%%%%%%%%%%%%%%%%%%%%%%%%%%%%%%%%%%%%%%%%%%%%%%%%%%%%%%%%%%%%%%%%%%%%%%%%%%%%%%

\section{Methods}
\label{sec:Methods}

Though the aim of the data reduction process is to minimise the correlated noise, the atmospheric signal remains buried in noise. To extract it, we employ the use of synthetic spectra, cross-correlating them with the reduced data. We also used these synthetic spectra to investigate limits to our detection capabilities for the atmospheric properties of WASP-76~b.

\subsection{Models} \label{ModelPrep}

The models we used to analyse our reduced data were created using the \texttt{\texttt{petitRADTRANS}} python package \citep{Molliere2019,Molliere2020}. \texttt{\texttt{petitRADTRANS}} can be used to generate both emission and transmission spectra, at either low resolution ($\lambda/\Delta\lambda \leq$ 1000) or high resolution ($\lambda/\Delta\lambda$ = 10$^6$). The modelled spectra can be produced for both clear and cloudy atmospheres. To make transmission spectra, it assumes the temperature and abundance profiles to be one-dimensional (1-D) and to describe the vertical structure of the entire planet. It also uses some properties of the star-planet system for which the adopted values are given in Table \ref{tab:sysparams}. When creating the spectra, we set up an atmosphere with 130 pressure layers going from 10$^{-8}$ to 10$^2$ bars (each being equidistant in log-space), with a reference pressure of P$_0$~=~0.1~bar defining the base of our atmosphere. We included the Rayleigh scattering cross-sections of H$_2$ and He, as well as the collision induced absorption cross-sections for the H$_2$-H$_2$ and H$_2$-He pairs. Clouds are included as a grey cloud deck at a given pressure. Assuming a solar H/He ratio, we also used molecular line lists of H$_2$O \citep{Polyansky2018}, CO \citep{Rothman2010,Kurucz1993}, C$_2$H$_2$ \citep{Rothman2013}, HCN \citep{Harris2006,Barber2014} and OH \citep{Rothman2010}. Abundances in \texttt{\texttt{petitRADTRANS}} are in Mass Mixing Ratios (MMRs). For all models used for our analysis, we assumed a vertically isothermal profile, as well as a uniform composition for the atmosphere.

Before using these models to analyse our reduced data, we wanted to bridge the gap between our model spectra and the real signal in terms of resemblance. First, ultra-hot Jupiters are expected to have circular orbits and be tidally locked (see a review in \citealp{Baraffe2010}), assumptions that lead to an estimated rotation speed of $\sim$ 5200 m.s$^{-1}$ at the equator of WASP-76~b. Rotational broadening will thus have a non-negligible effect on the spectral features. To take into account the effect of rotation, we used the double convolution framework presented in \citet{Klein2024}, allowing for a statistical exploration of rotation speed in the following sections. Other sources of broadening are the instrumental precision and exposure time, taken into account by convolving the signal with respectively a Gaussian function and a window function that averages the planet velocity over its $\sim$ 6200 m.s$^{-1}$ motion during the $\sim$ 774 s integration time of each  exposure. After broadening our \texttt{\texttt{petitRADTRANS}} model spectra, we normalised them. This is to replicate the normalisation of the real data during the reduction process described in section \ref{DataRed}.  
 
We also needed to take into account the effects from the orbital motion of WASP-76~b on the spectral features. For this, we first modulated the spectra by the transit window shown in Figure \ref{fig:propsofWASP-76}, to recreate the variations in intensity as the planet passes in front of its star. We then Doppler-shifted the modelled spectra by the planet radial velocity, V$_{\rm p}$, calculated for each phase $\phi$ of the observation (with $\phi$ itself calculated using the mid-transit time, the time vector of each exposure and the orbital period). As we kept the observed data in the geocentric frame, we calculated V$_{\rm p}$ in the same frame considering the orbit of WASP-76~b to be circular and by taking into account the barycentric Earth RV (BERV), and the stellar RV:
\begin{equation} \label{eq:planetvelocity}
    V_\mathrm{p}(\phi) = K_\mathrm{p} sin(2\pi\phi) + V_\mathrm{0} + V_\mathrm{c} - \mathrm{BERV}
\end{equation}
with K$_{\rm p}$ being the planet's radial velocity semi-amplitude, V$_\mathrm{0}$ the planet's systematic Doppler shift and V$_\mathrm{c}$ the stellar radial velocity.

As mentioned in section \ref{DataRed}, the models also needed to be degraded coherently with the the degradation of the signal due to the PCA to accurately represent the real atmospheric signal present in our reduced data. We followed the method of \cite{Gibson2022}, our final modified degraded model M' being calculated for each phase as follows:
\begin{equation}
    M' = \mathrm{exp}(\mathrm{log}M - UU^{\dag}\mathrm{log}M)
\end{equation}
with M and U corresponding to respectively our non-degraded model and the matrix stored during the data reduction process (U$^{\dag}$ being its pseudo-inverse). 

\begin{table*}
    \centering
    \caption{Priors used in our Nested Sampling algorithm through this paper, and posteriors found in sections \ref{H2O&COdetectest} (Post. 1), \ref{HCN&C2H2} (Post. 2), \ref{resultsOH} (Post.~3), \ref{dynamics} (Post. 4), and \ref{sssec:C/O robustness} (Post. 5). }
    \label{tab:priors}
    \begin{threeparttable}
    {\renewcommand{\arraystretch}{1.5}%
    \begin{tabular}{lcccccr} % four columns, alignment for each
        \hline
        Parameter & Prior & Post. 1 & Post. 2 & Post. 3 & Post. 4 & Post. 5\\
        \hline
        K$_{\rm p}$ (km.s$^{-1}$) & $\mathcal{U}$(100,300) & 179.8$^{+10.40}_{-9.05}$ & 177.6$^{+9.42}_{-6.31}$ & 177.4$^{+9.53}_{-7.46}$ & 179.7$^{+5.97}_{-7.09}$ & 180.5$^{+10.80}_{-8.70}$ \\
        V$_{\rm 0}$ (km.s$^{-1}$) & $\mathcal{U}$(-30,10) & -6.18$^{+0.79}_{-1.60}$ & -5.96$^{+0.73}_{-1.28}$ & -5.84$^{+0.64}_{-1.50}$ & -5.97$^{+0.76}_{-0.98}$ & -6.08$^{+0.65}_{-1.80}$ \\
        log(H$_2$O) & $\mathcal{U}$(-8,-1) & -4.53$^{+1.34}_{-0.22}$ & -4.58$^{+0.20}_{-0.23}$ & -4.56$^{+0.18}_{-0.27}$ & -4.71$^{+0.20}_{-0.22}$ & - \\
        log(CO) & $\mathcal{U}$(-8,-1) & -3.10$^{+0.89}_{-1.21}$ & -3.28$^{+0.54}_{-1.05}$ & -3.02$^{+0.22}_{-1.22}$ & -3.38$^{+0.46}_{-1.20}$ & - \\
        log(C$_2$H$_2$) & $\mathcal{U}$(-12,-1) & - & < -5.0 & - & - & - \\
        log(HCN) & $\mathcal{U}$(-12,-1) & - & < -5.5 & - & - & - \\
        log(OH) & $\mathcal{U}$(-12,-1) & - & - & < -6.0 & - & - \\
        T$_\mathrm{iso}$ (K) & $\mathcal{U}$(500,5000) & 1355$^{+305}_{-321}$ & \textit{fixed} & \textit{fixed} & \textit{fixed} & 1356$^{+259}_{-342}$ \\
        P$_\mathrm{cloud}$ (bar) & $\mathcal{U}$(-8,2) & 1.08$^{+0.30}_{-2.84}$ & - & - & - & 0.82$^{+0.51}_{-2.89}$ \\
        v$_\mathrm{rot}$ (m.s$^{-1}$) & $\mathcal{U}$(100,10000) & \textit{fixed} & \textit{fixed} & \textit{fixed} & 3544$^{+1108}_{-1994}$ & \textit{fixed} \\
        C/O & $\mathcal{U}$(0,1) & - & - & - & - & 0.90$^{+0.01}_{-0.51}$\\
        $\left[\frac{\rm (C+O)}{\rm H}\right]$ & $\mathcal{U}$(-6,1.4) & - & - & - & - & -1.47$^{+1.42}_{-0.47}$ \\
        \hline

    \end{tabular}}
    
    \begin{tablenotes}
    \item \footnotesize{\textbf{Notes -} The abundances are in MMR, we use "log" for decimal  logarithms and  $\mathcal{U}$ stands for a uniform distribution. The symbol - indicates that the parameter was not considered in the input parameters of the corresponding retrieval. The fixed values of T$_\mathrm{iso}$ and v$_\mathrm{rot}$ are respectively 1500 K and 5210 m.s$^{-1}$.}
    \end{tablenotes}

\end{threeparttable}
\end{table*}

\subsection{Analysis}
\subsubsection{Cross-Correlation Maps} \label{CCmaps_section}

Cross-correlation maps are made by cross-correlating the reduced data with the modelled spectra, over a large range of K$_{\rm p}$ and V$_{\rm 0}$ values used in equation \ref{eq:planetvelocity}. Such maps allow the detection of an atmospheric signal by finding a maximal correlation value around the physically expected K$_{\rm p}$ and V$_{\rm 0}$ values for the planet. The former can be calculated using the mass values of the star and planet, and the semi-amplitude of the stellar RV. The latter is informed by previous studies of WASP-76~b's atmospheric signal such as \cite{Ehren2020} and priors for atmospheric dynamics in exoplanet atmospheres. The correlation function used is defined as in \cite{Boucher2021}, as follows:
\begin{equation}
    CCF = \sum_{i} \frac{d_{i}m_{i}}{\sigma_{i}^{2}}
\end{equation}
where d$_i$ represents the observed flux, m$_i$ the modelled spectra, and $\sigma_i$ the flux uncertainty, with the index i corresponding to the pixel at time t and wavelength $\lambda$. We calculated the cross-correlation maps for values of K$_{\rm p}$ going from 0 to 300 km.s$^{-1}$ with a step of 2 km.s$^{-1}$, and of V$_{\rm 0}$ going from -100 to +100 km.s$^{-1}$ with a step of 0.5 km.s$^{-1}$. The resulting correlation values are converted to significance of detection by dividing the former by the standard deviation of the areas considered to be dominated by white noise. We used theoretical values from \cite{Ehren2020} for both K$_\mathrm{p}$ and V$_\mathrm{0}$ to mark approximately where the correlation is expected to peak within the map if the atmospheric signal is detected. As the K$_\mathrm{p}$ measured seems dependent on species (see Table 2 in \citealt{Kesseli2022} for example), we chose to use K$_\mathrm{p}$~=~196.52~km.s$^{-1}$ (estimated neglecting atmospheric dynamics), though we expect to see a shift of up to approximately $\Delta$K$_\mathrm{p}$~=~$\pm$20 km.s$^{-1}$ \citep{Wardenier2023}. For the Doppler shift, we used V$_\mathrm{0}$ = -5.3~km.s$^{-1}$, a value assumed for the day-to-night wind velocity in \cite{Ehren2020} to compensate the redshift due to planet rotation for the morning limb.

\subsubsection{Atmospheric retrieval} \label{NS_section}

To search for the best-fit values of our parameters, we used two methods. The first was a Markov Chain Monte Carlo (MCMC) algorithm using the emcee python module \citep{ForemanMackey2013}. The second was a nested sampling (NS) algorithm using the python module pymultinest \citep{Buchner2014}. As in \citet{Klein2024}, we found the latter to be around 50 times faster than the former while giving very similar results. For these reasons, we only present the results from the NS algorithm. Our choice of priors for this whole paper are summed up in Table \ref{tab:priors}.

For our parameter search, we used the likelihood $\mathcal{L}$ defined in \cite{Gibson2020} by:
\begin{equation}
    \mathrm{ln} \mathcal{L} = -N\mathrm{ln}\beta - \frac{1}{2} \sum_{i=1}^{N} \frac{(f_i - \alpha m_i)^2}{\beta \sigma_i^2} 
\end{equation}
where $\alpha$ and $\beta$ are scaling factors that take into account scale uncertainties of respectively the model and the white noise. Tests of leaving $\alpha$ as a free parameter for models containing only H$_2$O opacity lines showed a preference for $\alpha$ = 1 and a slight degeneracy with temperature (as $\alpha$ increased, the temperature decreased), coming from the fact that both parameters influence the scale height. We chose for the rest of our study to set $\alpha$ = 1, like \cite{BrogiLine2019}, which implies that the scale of the model is the same as the analysed data. When including $\beta$ within our free parameter set, we found very similar results as when setting $\beta$ = 1. As the main difference between the former and the latter was computational cost, with the latter being around 2 $\times$ faster than the former, we chose to set $\beta$ = 1.

\subsubsection{Data Simulator} \label{Data_simulator}

We also used the \texttt{\texttt{petitRADTRANS}} generated spectra to simulate SPIRou observations of the atmospheric signal of WASP-76~b. The models were created and broadened following the same procedure as described in section \ref{ModelPrep}. They were then injected into our real data at a negative K$_{\rm p}$ value so as to not confuse our simulated signal with the real signal. We also chose to inject them at V$_{\rm 0}$ different from that expected for the real signal. The noise level for our simulated data was hence the same as that for the real atmospheric signal.

The data containing our simulated spectra was then put through the data reduction pipeline described in section \ref{DataRed}, and analysed through the use of cross-correlation maps as described in section \ref{CCmaps_section}. Unlike for the real atmospheric signal, we know the exact input parameters of the atmospheric signal that we are searching for. Thus, a non-detection of the simulated signal would indicate a too high noise level to allow a detection. Hence, we can investigate the required abundances of different species that could be in the atmosphere for a detection to be possible.

\subsection{Forward models}

We used the 1-D radiative-convective chemical equilibrium code \texttt{\texttt{ATMO}} \citep{Tremblin2015, Amundsen2014,Drummond2019} to create forward models of the atmosphere of WASP-76 b. Like for creating the \texttt{\texttt{petitRADTRANS}} model spectra, we set the number of pressure layers to 130, going from around 10$^{-8}$ to 10$^{2}$ bar. The pressure-temperature profiles for these models are calculated by \texttt{\texttt{ATMO}} to satisfy both hydrostatic equilibrium and conservation of energy. Physically consistent chemical abundance profiles of 175 gaseous species are also calculated by the code in equilibrium from solar abundances of 23 elements (H, He, C, N, O, Na, K, Si, Ar, Ti, V, S, Cl, Mg, Al, Ca, Fe, Cr, Li, Cs, Rb, F and P, with abundances adopted from \citealt{Caffau2011}). As we could change the C/O ratio of the modelled atmosphere by modifying the initial quantities of C and/or O, we created models with the same C/O ratio obtained for different initial abundances of C and O. We first used these forward models to predict which species were the most likely to be present in the atmosphere and detectable with SPIRou. We then later used these same forward models to evaluate if the (non-)detections of these species and the abundance values or upper limits retrieved with the NS algorithm were coherent with the estimated C/O ratio.

%%%%%%%%%%%%%%%%%%%%%%%%%%%%%%%%%%%%%%%%%%%%%%%%%%%%%%%%%%%%%%%%%%%%%%%%%%%%%%%
%%%%%%%%%%%%%%%%%%%%%%%%%%%%%%%%%%%%%%%%%%%%%%%%%%%%%%%%%%%%%%%%%%%%%%%%%%%%%%%

\section{Results}
\label{sec:Results}

Using \texttt{\texttt{ATMO}} forward models, we estimated that depending on the C/O ratio, the most probable species to be present in the atmosphere that can be detected with SPIRou are H$_2$O, CO, HCN, C$_2$H$_2$ and OH. This section therefore presents our attempts to detect them as well as the associated temperature, cloud and dynamical profiles recovered.

\subsection{H$_2$O and CO} \label{H2O&COdetectest}

\begin{figure}
    \includegraphics[width=\linewidth]{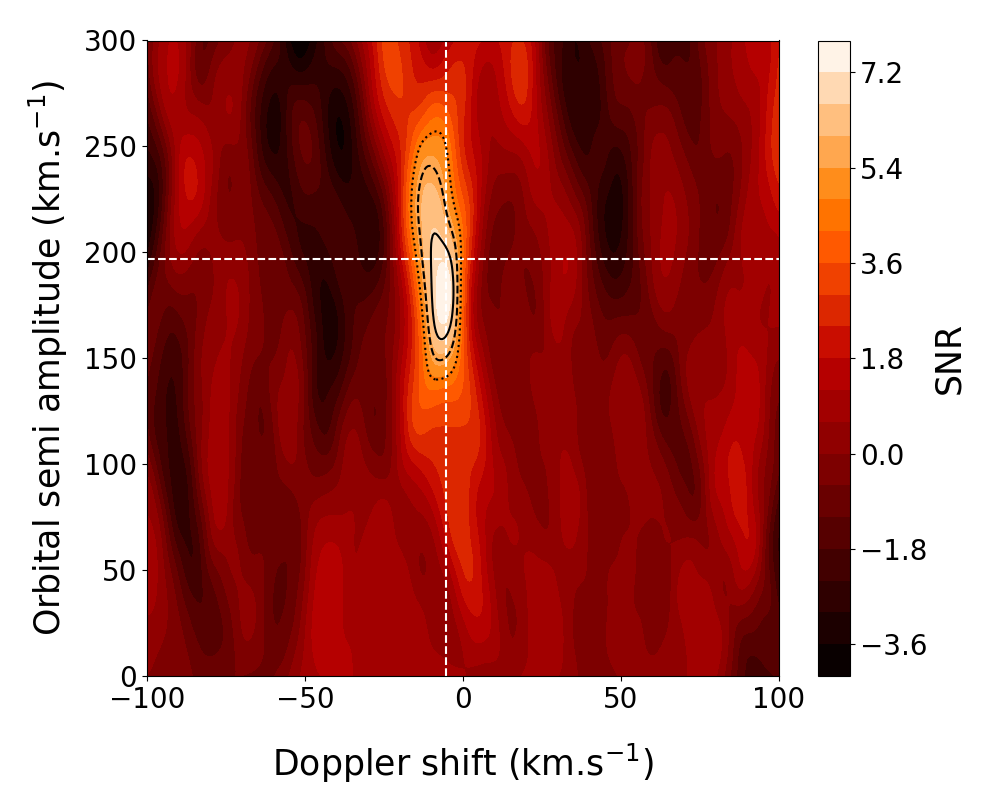}
    \caption{Map resulting from cross-correlation between the reduced data of WASP-76~b and spectra from a model containing H$_2$O opacity lines. The SNR varies from -3.86 to 7.29, with the maximum peak in SNR of 7.29 obtained for K$_\mathrm{p}$ = 180 km.s$^{-1}$ and V$_\mathrm{0}$ = -6.0 km.s$^{-1}$. The black lines (full, dashed and dotted) respectively represent the 1, 2 and 3 $\sigma$ contours. The white dashed lines represent the theoretical values from \citet{Ehren2020} of K$_\mathrm{p}$ = 196.52 km.s$^{-1}$ and V$_\mathrm{0}$ = -5.3 km.s$^{-1}$. }
    \label{fig:H2Odetection}
\end{figure}

To first detect WASP-76~b's atmosphere, we used models created following the procedure described in section \ref{ModelPrep} including only water opacity lines, with log(H$_2$O)$_\mathrm{MMR}$ = -5.0. We used an isothermal temperature profile with T = 1500 K, a choice motivated by the retrieval results obtained using our NS algorithm presented later in this section. The resulting cross-correlation map can be seen in Figure \ref{fig:H2Odetection}. The maximum SNR of 7.29 was obtained for K$_\mathrm{p}$ = 180~km.s$^{-1}$ and V$_\mathrm{0}$ = -6.0~km.s$^{-1}$. The intersection of the theoretical values from \citet{Ehren2020} are within the 1$\sigma$ boundary of our maximum SNR peak. Furthermore, \cite{Wardenier2021} showed that 3-D effects in the atmosphere of WASP-76~b such as atmospheric dynamics and rotation can lead to a shift in K$_{\rm p}$ up to 30~km.s$^{-1}$ lower than the expected value. This has previously been observed for many of the detected species reported in Table \ref{tab:chemlist_table}, with the measured shift changing depending on the species (e.g. \citealt{Kesseli2022,Deibert2023}). We were hence able to both confirm the detection of the atmosphere of WASP-76~b and the presence of H$_2$O in its atmosphere.

\begin{figure}
    \includegraphics[width=\linewidth]{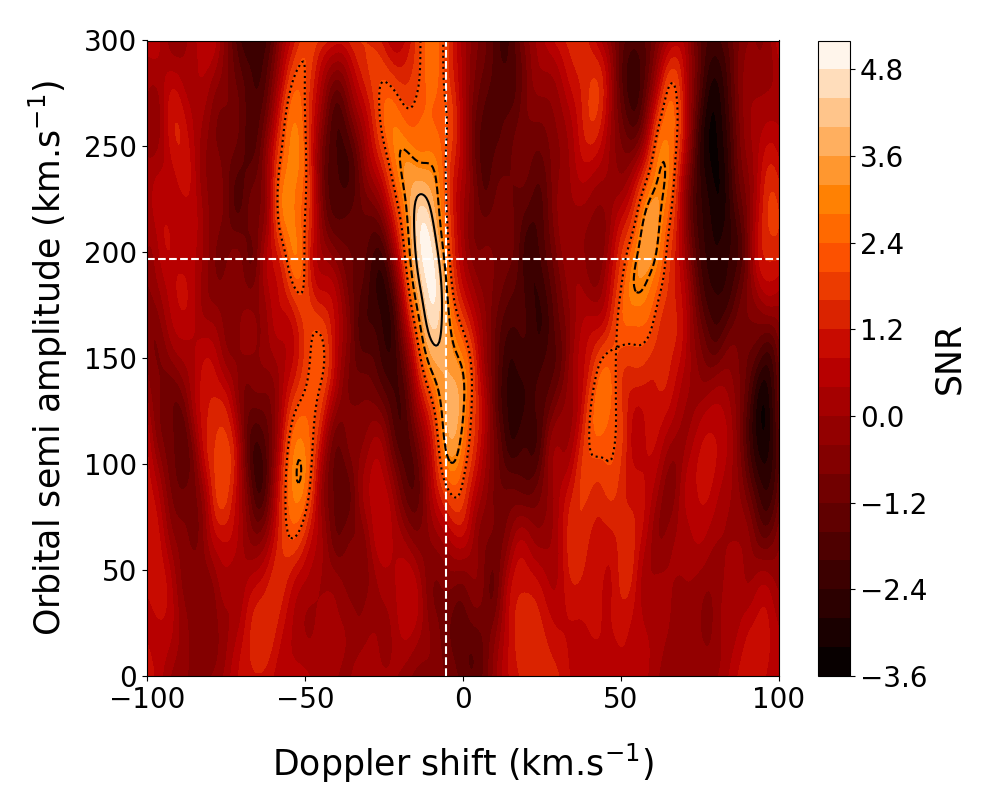}
    \caption{Same as Fig. \ref{fig:H2Odetection} but using models containing only CO. The SNR varies from -3.30 to 5.03. The maximum SNR was obtained for K$_\mathrm{p}$~=~200 km.s$^{-1}$ and V$_\mathrm{0}$ = -11.5 km.s$^{-1}$.}
    \label{fig:COdetectionbestorders}
\end{figure}

We next looked for an atmosphere assumed to only contain CO. The models were created as described in section \ref{ModelPrep}, with only CO opacity lines included (log(CO)$_\mathrm{MMR}$ = -4.0). The resulting cross-correlation map can be seen in Figure \ref{fig:COdetectionbestorders}, confirming as well a detection of CO in the atmosphere, dynamically consistent at 1$\sigma$ with the H$_2$O detection but only consistent at 2$\sigma$ with the values from \citet{Ehren2020}.

\begin{figure}
    \includegraphics[width=\linewidth,trim={0 0 0 0.1cm},clip]{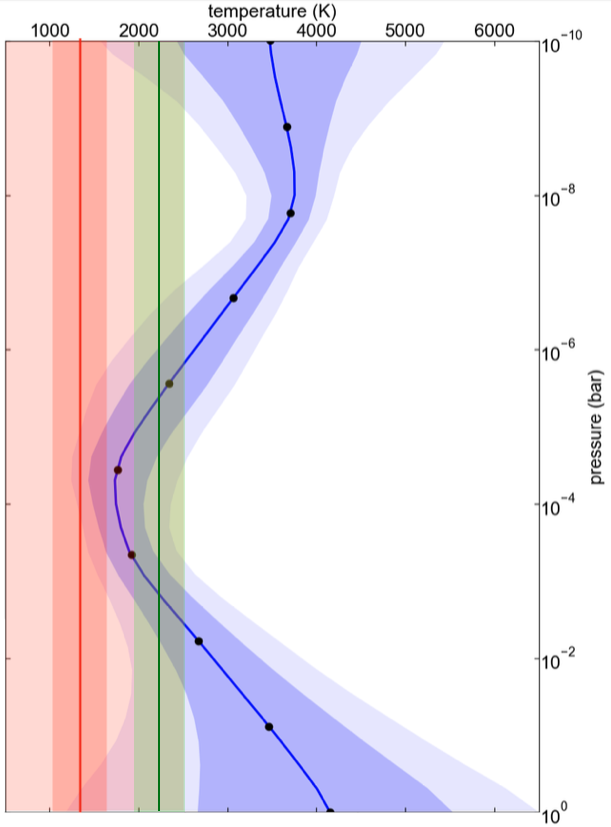}
    \caption{Temperature profile from Extended Data Figure 5 in \cite{Pelletier2023} (blue), compared to the isothermal temperature profile from Figure 7 in \cite{Edwards2020} (green) and the isothermal temperature profile retrieved in Figure \ref{fig:retrievalwithclouds} with T = 1355 K (red), with the shaded regions representing the 1 and 3 $\sigma$ error bars (only 1$\sigma$ represented for the profile from \citealt{Edwards2020}). }
    \label{fig:temperature_comparaison_Stefan_isothermal}
\end{figure}

We were then able to estimate abundances for both H$_2$O and CO using our NS algorithm. The output parameters were the orbital RV semi amplitude, the Doppler shift, the abundance of water and CO, temperature and the cloud top pressure as shown in Table \ref{tab:priors}. The presence of high-altitude clouds in the atmosphere can impact the absorption line depth and hence the measurement of chemical abundances. The resulting posterior distribution are presented in Appendix \ref{apen:retrievalwithclouds}. The maximum posterior probabilities give a temperature of 1355 $\pm$ 313 K, with 1$\sigma$ error bars, much colder than the mean temperature obtained by \citet{Pelletier2023} for transit observations in the visible ($\sim$ 3000 K). However, when comparing our isothermal temperature profile to the vertical temperature structure found by \citet{Pelletier2023} (see Figure \ref{fig:temperature_comparaison_Stefan_isothermal}), we can see that we are closest in temperature for a pressure of around 10$^{-4}$ bar. This is coherent with the higher sensitivity of SPIRou data to pressure levels going from around 10$^{-4}$ to 10$^{-2}$ bar. The MMR of water, -4.53 $\pm$ 0.77 (in log), is about ten times lower than that of CO (-3.10 $\pm$ 1.05). This translates into a C/O ratio of 0.94 $\pm$ 0.15 which is discussed further in section \ref{sec:Discussion}. The data favour either a deep cloud deck or, more likely, the absence of clouds, excluding clouds higher than 10$^{-4}$ bars at 3$\sigma$. We can see two expected degeneracies: temperature with abundances as well as cloud-top pressure and abundances. As we show in section \ref{sec:Discussion}, these degeneracies affect mostly the metallicity while keeping a C/O ratio roughly constant.

Because water and CO could be sensitive to different pressure levels in the atmosphere and that that would affect their recovered abundances when retrieving them jointly, we also ran NS algorithms with only each of these molecules included. The retrieval of water is globally unchanged, while we find a higher quantity of CO (-2.74 $\pm$ 0.95) linked to a temperature 300 K higher (not shown). This further confirms a high C/O ratio (section \ref{sec:Discussion}).

As water is detected more strongly than CO, it also has a higher influence on the likelihood. In that regard, while we were able to obtain a lower limit for water, the data are consistent with the absence of CO at 3$\sigma$. However, we know that CO is present in the atmosphere, as shown by our detection in Fig.\ref{fig:COdetectionbestorders} and other studies. Hence, this acts as a bias towards lower abundances of CO, and by extension, lower recovered C/O values in the atmosphere.

\subsection{HCN and C$_2$H$_2$} \label{HCN&C2H2}

Other species we estimated to potentially be present in the atmosphere of WASP-76~b are HCN and C$_2$H$_2$. However, we were unable to detect either of these species when analysing the data of WASP-76~b. Nevertheless, we were able to infer upper limits of their abundances from the results of the atmospheric retrieval (see Figure 1 in Appendix \ref{C}), with these roughly indicating log(C$_2$H$_2$)$_\mathrm{MMR}$ < -5.0 and log(HCN)$_\mathrm{MMR}$ < -5.5. 

We used the data simulator (see section \ref{Data_simulator}) to verify the limitations of our detection for these two species in our data. The simulated atmospheric signals used contained both water (with log(H$_2$O)$_\mathrm{MMR}$ = -5.0) and either HCN or C$_2$H$_2$ with different MMRs. Each was analysed using a model containing only the species included in the simulated signal, either HCN or C$_2$H$_2$, with the same MMR as that used to create the signal. We found that for C$_2$H$_2$, a MMR of $10^{-6}$ was required for any indications of a detection, while for HCN, a MMR of $10^{-4}$ was needed. This is somewhat in agreement with the upper limits previously inferred from the nested sampling algorithm results. We further discuss these results in section \ref{upperlimitsHCN&C2H2}.

\subsection{OH} \label{resultsOH}

OH has been detected in the atmosphere of WASP-76~b \citep{Landman2021,Mansfield2022}. However, we were not able to detect it in our data. An upper limit for the abundance was indicated by the NS algorithm results, roughly indicating log(OH)$_\mathrm{MMR}$ < -6 (see Figure 2 in Appendix \ref{C}). With the data simulator, we were able to infer that a detection would be possible for log(OH) $\gtrsim$ -4. We also used the data simulator with the telluric-corrected data from the \texttt{\texttt{APERO}} version 0.7.275 to investigate the detectability of OH, as this later version of the SPIRou DRS is supposed to have an improved correction for OH signal coming from looking through the Earth's atmosphere. However, we also found that a log(OH)$_\mathrm{MMR}$ of at least -4 was required for a detection to be possible, finding the same detection limit for OH as found for the data from version 0.6.132 of \texttt{\texttt{APERO}}.

\subsection{Dynamics} \label{dynamics}

To investigate the rotational velocity of the planet, we used cloud-free models for which we fixed the temperature to T~=~1500~K and included both H$_2$O and CO opacity line lists. The NS algorithm results can be seen in Appendix \ref{apen:retrievalwithvrot}. The K$\rm_p$ and V$_\mathrm{0}$ values found are in good agreement with those found in section \ref{H2O&COdetectest}, being within 1$\sigma$ of those values. The abundances found for both water and carbon monoxide are lower than those previously found in section \ref{H2O&COdetectest}, but consistent at <1$\sigma$. The rotation velocity found of v$_{\rm eq}$ = 3544 $\pm$ 1501 m.s$^{-1}$ is lower than that expected for WASP-76~b, estimated to be v$_{\rm eq, theoretical} \simeq$ 5210 m.s$^{-1}$) by considering the planet as tidally-locked and with a circular orbit. However, the expected velocity is within the 2$\sigma$ boundaries (and actually very close to 1$\sigma$).

%%%%%%%%%%%%%%%%%%%%%%%%%%%%%%%%%%%%%%%%%%%%%%%%%%%%%%%%%%%%%%%%%%%%%%%%%%%%%%%
%%%%%%%%%%%%%%%%%%%%%%%%%%%%%%%%%%%%%%%%%%%%%%%%%%%%%%%%%%%%%%%%%%%%%%%%%%%%%%%

\begin{figure}
    \includegraphics[width=\linewidth]{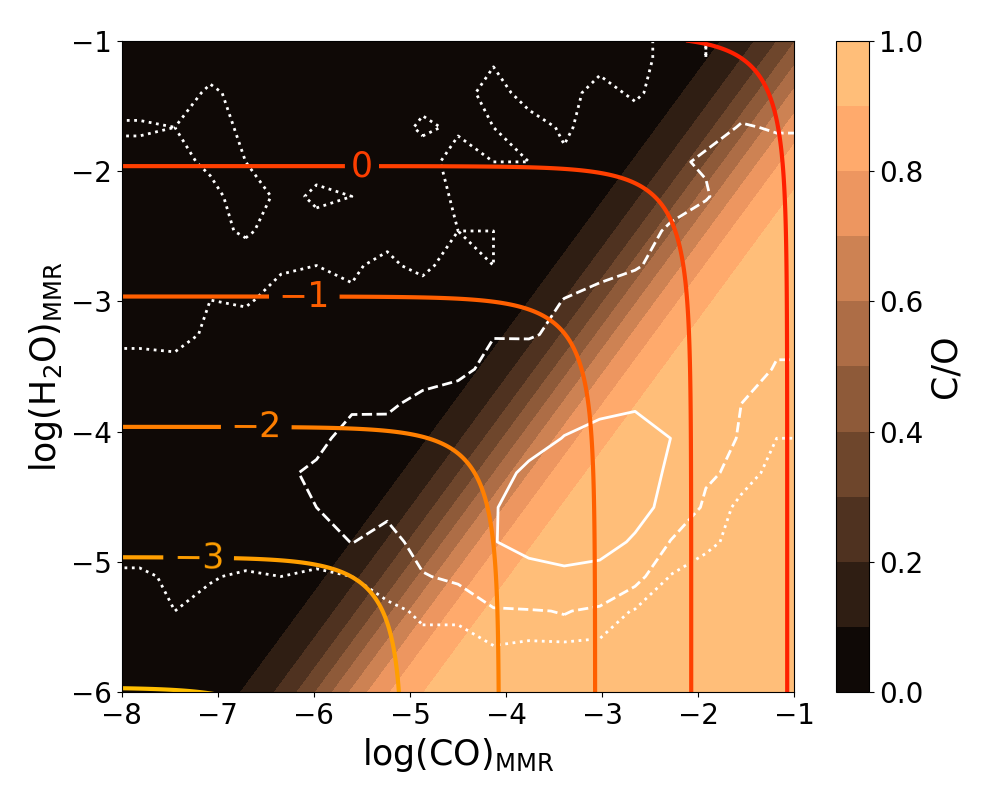}
    \caption{Representation of the possible C/O and [(C+O)/H] values for the posteriors of log(H2O) and log(CO). The white lines (full, dashed and dotted) respectively represent the 1, 2 and 3$\sigma$ contours shown in Figure \ref{fig:retrievalwithclouds}. For the whole range of possible values of log(H$_2$O)$_{\rm MMR}$ and log(CO)$_{\rm MMR}$, we calculated the corresponding values of the C/O and [(C+O)/H] ratios. The background represents the resulting values of C/O ratio. The coloured lines indicate the values of [(C+O)/H].}
    \label{fig:CObackground_CpOlines}
\end{figure}

\section{Discussion}
\label{sec:Discussion}
\subsection{C/O ratio}
\subsubsection{Estimation} \label{COrat_estimation}

We could not detect any C- or O-bearers other than H$_2$O and CO. Forward models subsequently created with \texttt{\texttt{ATMO}} for different C/O ratios values showed that water and carbon monoxide are the main bearers of these elements. We hence concluded that the abundances retrieved for H$_2$O and CO found in section \ref{H2O&COdetectest} to be sufficient to get a first approximation for the C/O ratio local to the section of WASP-76~b's atmosphere observed in our data. To do so, we first converted the MMR abundance values to Volume Mixing Ratios (VMRs). The C/O ratio was then calculated as in \cite{Line2021}, which is as follows:
\begin{equation} \label{equationCOrat}
    C/O = \frac{n_\mathrm{CO}}{n_\mathrm{CO}+n_\mathrm{H_2O}}
\end{equation}
with n$_\mathrm{CO}$ and n$_\mathrm{H_2O}$ representing respectively the VMRs of CO and H$_2$O. Using the most likely values for the abundances of water and CO found by our NS algorithm in section \ref{H2O&COdetectest}, we estimated a C/O ratio of approximately 0.94 $\pm$ 0.39 ($\sim$ 1.7 $\pm$ 0.7 $\times$ solar, with solar abundances taken from \citealt{Asplund2009} and errors indicated being for 2$\sigma$ values) for the part of the atmosphere probed by the SPIRou transmission data used for this study. The C/O ratios estimated using the retrieval results associated to sections \ref{HCN&C2H2}, \ref{resultsOH} and \ref{dynamics} are all within 1$\sigma$ of this value. Indeed, the inclusion of HCN and C$_2$H$_2$ leads to a C/O ratio of 0.93~$\pm$~0.35, the inclusion of OH to a C/O ratio of 0.92~$\pm$~0.21, and the use of cloud-free models to a C/O ratio of 0.93~$\pm$~0.34 (with errors indicated corresponding to the 2$\sigma$ values). Furthermore, we compared the sigma contours for the posterior values of [H$_2$O, CO] to the C/O ratios calculated for the prior ranges of H$_2$O and CO (see Figure \ref{fig:CObackground_CpOlines}). For values within the 1$\sigma$ boundary, we found C/O to be greater than 0.6 ($\sim$ 1.1 $\times$ solar). The other sigma contours however encompass all possible C/O values. This seems to be a direct consequence of the lack of a lower boundary for the abundance of CO. However, due to only including CO as a C-bearing species, equation \ref{equationCOrat} is naturally limited to vary between 0 and 1. This is further explored in the next subsections.

\begin{figure}
    \includegraphics[width=\linewidth]{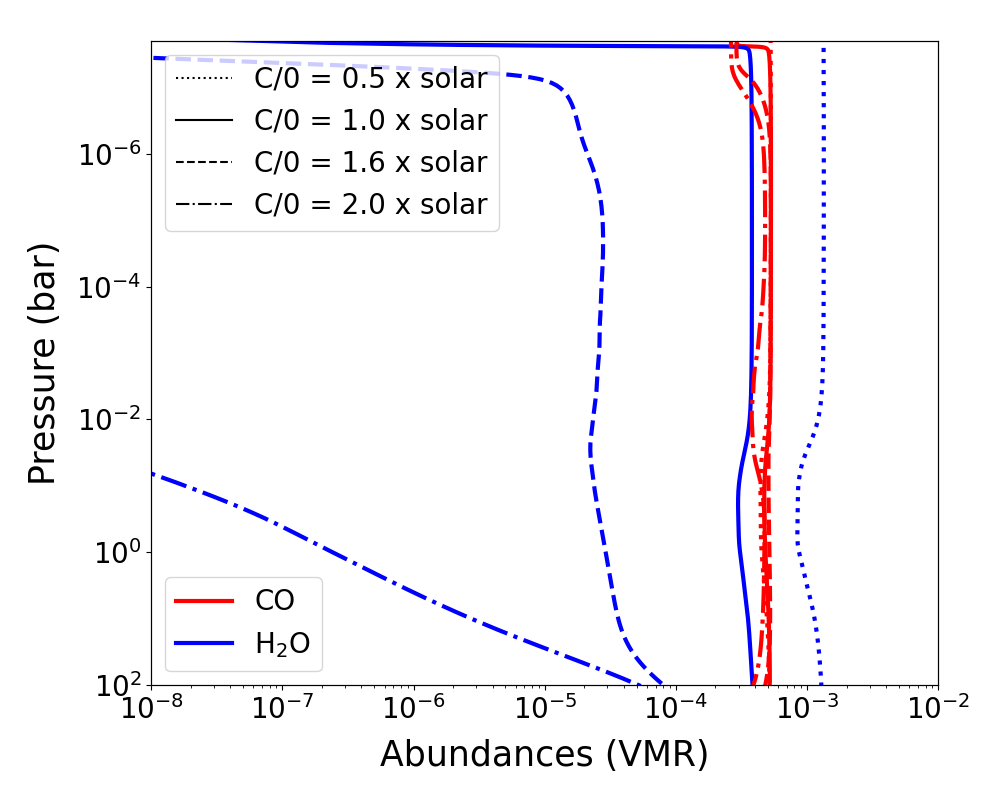}
    \caption{Variations in VMR of H$_2$O and CO depending on pressure calculated by \texttt{\texttt{ATMO}} for C/O = 0.5, 1, 1.6 and 2 $\times$ solar.}
    \label{fig:H2OvsCO_atmo}
\end{figure}

\subsubsection{Robustness of a super-solar C/O ratio}
\label{sssec:C/O robustness}
We used the 1-D radiative-convective code \texttt{\texttt{ATMO}} \citep{Tremblin2015} to investigate the expected abundance profiles of H$_2$O and CO depending on the overall C/O ratio of the atmosphere. Note that unlike our C/O calculation with equation \ref{equationCOrat} that only uses abundances of H$_2$O and CO, the C/O ratios of \texttt{\texttt{ATMO}} include all C- and O-bearing species included in the model. The results for C/O = 0.5, 1.0, 1.6 and 2.0 ($\times$ solar) can be seen in Figure \ref{fig:H2OvsCO_atmo}. From these results, we can infer that if C/O < 1 $\times$ solar, we would find log(H$_2$O) > log(CO), while if C/O $\sim$ 1 $\times$ solar, the measured abundances of H$_2$O and CO would be of similar magnitude. We can also infer that if C/O was greater than 2 $\times$ solar, then there would be several orders of magnitude between the abundances of water and CO. Hence, considering these results and the difference in order of magnitude found between the retrieved abundances of H$_2$O and CO in section \ref{H2O&COdetectest}, we would expect to estimate a C/O ratio greater than 1 $\times$ solar but smaller than 2 $\times$ solar. This is consistent with our the C/O ratio estimated from our results, for which we assumed only H$_2$O and CO present in the atmosphere. 

%\ref{apen:retrievalwithCOrat&metal}
Furthermore, it is shown in \cite{Line2013} that when using uniform or Gaussian priors for the abundances of the C and O bearing species, the corresponding calculated prior C/O ratio consists of two peaks, with a preference for a C/O $\sim$ 1. To test for a possible bias on our estimated C/O ratio, we also used as uniform priors of C/O and [(C+O)/H] (a proxy for metallicity, with '[]' referring to the log$_{10}$ of the value relative to the solar value). As we consider once again only H$_2$O and CO in the atmosphere, we impose an upper limit on possible C/O ratios of~1. The calculations of the corresponding abundances of H$_2$O and CO are shown in Appendix \ref{changing priors}. The results obtained from the NS algorithm are shown in Appendix \ref{apen:retrievalwithCOrat&metal}. The maximum probability for the distribution of C/O is obtained for a value of 0.90$_{-0.78}^{+0.07}$ (with errors indicated here corresponding to 2$\sigma$ values), which is roughly the same as the C/O ratio found using uniform distributions of H$_2$O and CO as priors. We can see that similar degeneracies as those for water also exist for our metallicity proxy [(C+O)/H], being degenerate with temperature and cloud pressure . Meanwhile, the C/O ratio seems to remain relatively constant, though with a large lower boundary. This is consistent with what is shown by Figure \ref{fig:CObackground_CpOlines}, with the large lower boundary on our estimation of C/O coming from the large lower boundary on the abundance of CO.

\subsubsection{Upper limits for HCN and C$_2$H$_2$} \label{upperlimitsHCN&C2H2}

\begin{figure}
    \includegraphics[width=\linewidth]{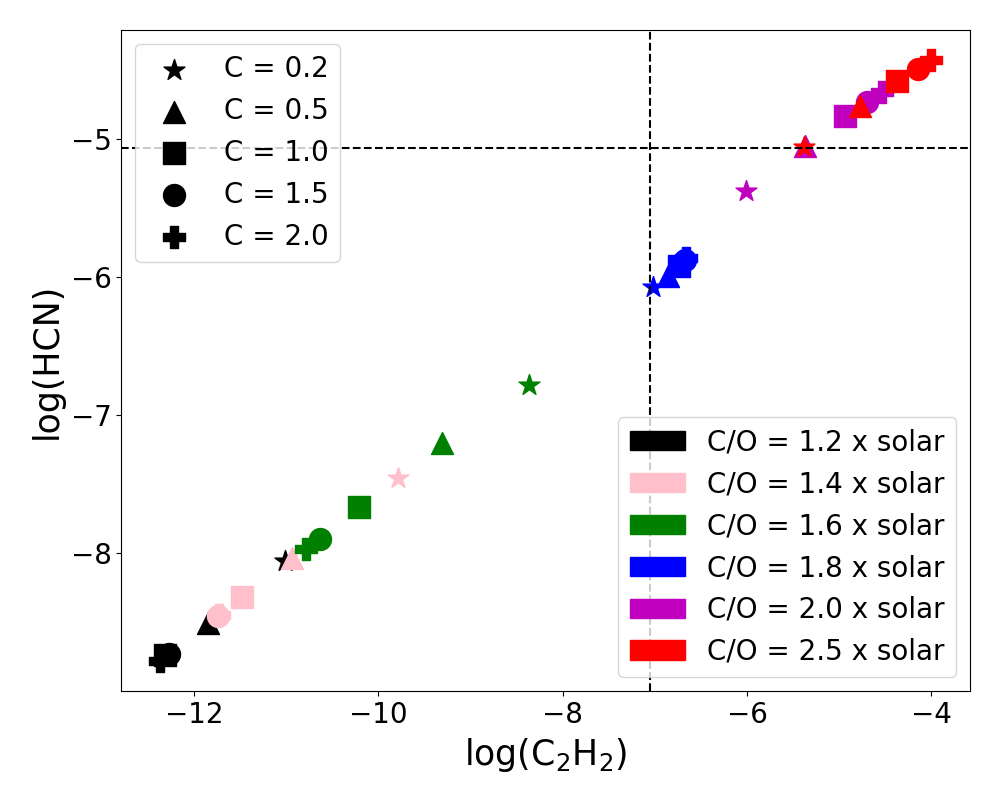}
    \caption{Mean abundances in VMR of C$_2$H$_2$ and HCN (averaged over pressures from 10$^0$ to 10$^{-4}$ bar). The black dashed lines correspond to the upper limit abundances found by the data simulator of log(C$_2$H$_2$)$_{\rm MMR}$=-6 and log(HCN)$_{\rm MMR}$=-4.}
    \label{fig:HCN&C2H2_diffCOratvaluesexpected}
\end{figure}

With ATMO, we calculated the abundance profiles of different species in thermochemical equilibrium. We did so for a variety of C/O ratios, obtained by changing the initial abundances of both C and O so as to also change the metallicity \citep{Drummond2019}. We estimated the expected quantity of HCN and C$_2$H$_2$ dependant on the C/O ratio (see Figure \ref{fig:HCN&C2H2_diffCOratvaluesexpected}). We compared the results to the upper limits found for these species to be detectable with the data simulator. Though the upper limit of HCN only excluded C/O ratio over 2.5 times solar,  the upper limit of C$_2$H$_2$ excludes all C/O ratios that are above 1.8 $\times$ solar unless the metallicity is significantly sub-solar (see next section). This is in agreement with what was indicated previously by Figure \ref{fig:H2OvsCO_atmo}. Furthermore,  C/O < 1.8 $\times$ solar is equivalent to C/O < 0.99, which is consistent with our C/O value estimation from section \ref{COrat_estimation}.

\subsubsection{OH} \label{discussOH}

\begin{figure}
    \includegraphics[width=\linewidth]{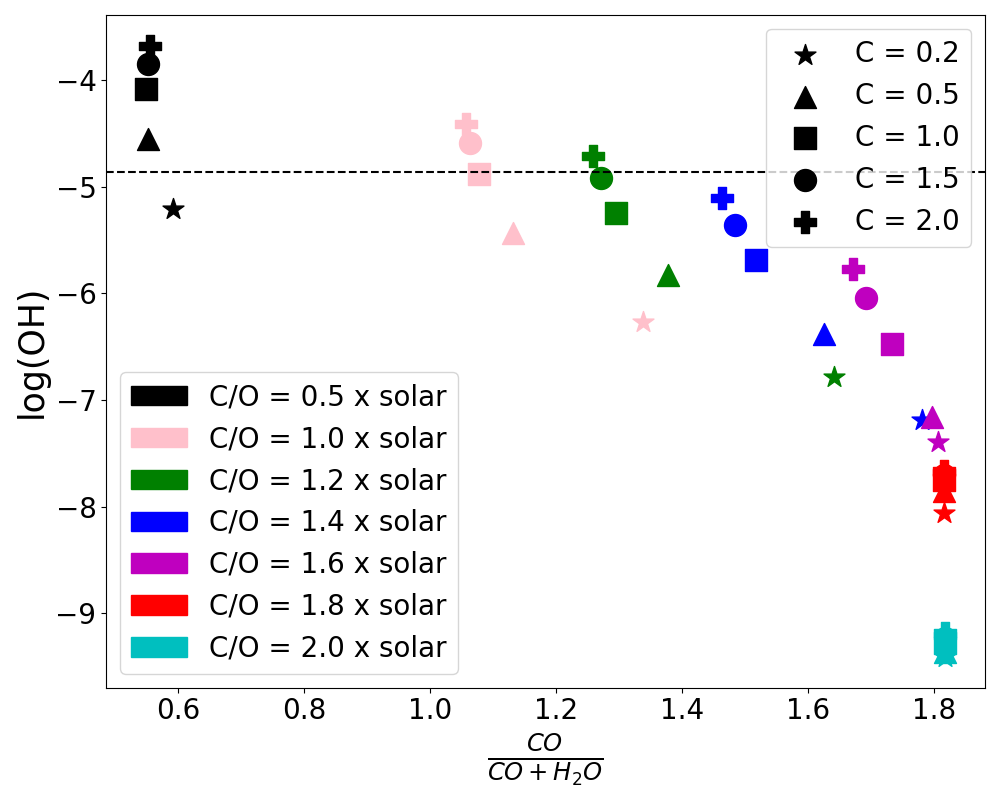}
    \caption{Mean abundances in VMR of OH (averaged over pressures from 10$^0$ to 10$^{-4}$ bar) as a function of the corresponding C/O ratio as estimated using equation \ref{equationCOrat}. The black dashed line corresponds to the upper limit abundance found by the data simulator of log(OH)$_{\rm MMR}$=~-4.}
    \label{fig:OH_diffCOratvaluesexpected}
\end{figure}

Using the aforementioned \texttt{\texttt{ATMO}} models, we investigated the impact of our non-detection of OH on the C/O ratio as calculated using equation \ref{equationCOrat}. In Figure \ref{fig:OH_diffCOratvaluesexpected}, we represent the mean abundance of OH for pressures probed by SPIRou as a function of the C/O ratio estimated using equation \ref{equationCOrat} for each of the \texttt{\texttt{ATMO}} models used. By comparing the results to the upper limit found using the data simulator, we can see that our non-detection of OH is also consistent with having a C/O ratio between 1 and 2 $\times$ solar for solar metallicity. 

\subsubsection{Implications on planet formation?}

Our different retrieval results all led to estimating the C/O ratio to be super-solar, finding C/O $\sim$ 0.9. This is in agreement with the super-solar C/O ratio found for transmission spectra acquired with HST by \cite{Fu2021}, of $\sim$ 0.83 $\pm$ 0.76. However, neither our retrieval results nor those from \cite{Fu2021} exclude entirely the sub-solar and solar C/O ratio cases. Furthermore, while the upper limits found for C$_2$H$_2$, HCN and OH seem to exclude the sub-solar C/O ratio case for solar metallicity, they do not exclude a solar C/O ratio for solar metallicity, or a sub-solar C/O ratio for a significantly sub-solar metallicity. Therefore, we cannot reasonably use our current results to place constraints on the formation scenario of this planet. By combining data from different transits in a future work, we should however obtain much better constraints on our abundances, and will then be able to also constrain the planet formation scenario of WASP-76 b.

\subsection{Metallicity and clouds} \label{metalclouds}
%\ref{apen:retrievalwithCOrat&metal}
Through our NS algorithm, we also looked at values calculated for [(C+O)/H], used as a proxy for metallicity (Fig.\ref{fig:CObackground_CpOlines} and Appendix \ref{apen:retrievalwithCOrat&metal}). Our data favour sub-solar metallicity at 1$\sigma$ while not excluding super-solar at 2$\sigma$, which prevents any solid conclusion on the metallicity. This is mainly due to the degeneracy with clouds, that prevents a robust estimation of the molecular mixing ratios. 

As stated in section \ref{ModelPrep}, we included clouds as a grey cloud deck at a given pressure when considering a cloudy atmosphere for our atmospheric retrievals. Previous studies have also included a grey cloud deck in the models used for retrievals. Using HST transmission data, \cite{Edwards2020} retrieve a cloud pressure of log(P$_{\rm clouds}$) = 0.91 $\pm$ 0.58 Pa (= -4.09 $\pm$ 0.58 bar). However, \cite{Pelletier2023} obtained using high-resolution optical transmission data from MAROON-X a cloud-top pressure for their optically thick grey cloud deck of log(P$_{\rm clouds}$) = -2.21 $\pm$ 0.28 bar. Neither of these values correspond to the one found to be most probable by our NS algorithm that favour a clear atmosphere, though neither are excluded either. Coupling low and high resolution spectroscopic data might lift this degeneracy and allow to conclude on the actual metallicity of the planet \cite{Boucher2023}.

\subsection{Dynamics}

\subsubsection{H$_2$O}

In section \ref{H2O&COdetectest}, we obtained for the cross correlation map of H$_2$O a Doppler shift of -~6.0~km.s$^{-1}$. While this value is inconsistent with the value previously reported for H$_2$O by \cite{SanchezLopez2022} (V$_{\rm 0} \sim$ -14.3 km.s$^{-1}$), it is consistent with both the value assumed for the day-to-night wind velocity from \cite{Ehren2020} and measured Doppler shifts for other species detected in the atmosphere of WASP-76 b (e.g. \citealt{Tabernero2021,Pelletier2023}). This blue-shift is hence compatible with a day-to-night wind of approximately the velocity measured, being V$_{\rm 0}$~=~-6.0 km.s$^{-1}$. As this corresponds to the wind speed retrieved for the whole transit, this could imply an equal contribution from both limbs to the overall signal of H$_2$O. However, with such a velocity having also been retrieved for species with asymmetrical signals like iron (e.g. \citealt{Tabernero2021,Pelletier2023}), we cannot conclude from the cross-correlation map obtained for the whole transit of WASP-76 b on the asymmetry of the water signal. No further indications are given by looking at the 2D cross correlation function map (see Appendix \ref{restframe_abssignals}).

\begin{table}
    \centering
    \caption{K$_{\rm p}$ and V$_{\rm 0}$ values for maximum SNR position obtained for water detection in each transit half.}
    \label{tab:maxKpVsysfortransithalves_H2Omodels}
    {\renewcommand{\arraystretch}{1.5}%
    \begin{tabular}{lcccr} % four columns, alignment for each
        \hline
        Parameter & & Half 1 & & Half 2 \\
        \hline
        K$_{\rm p}$ (km.s$^{-1}$) & & 184$^{+80}_{-27}$ & & 216$^{+35}_{-34}$ \\
        V$_{\rm 0}$ (km.s$^{-1}$)& & -4.5$^{+3.7}_{-4.6}$ & & -11.5$^{+4.5}_{-4.4}$   \\

        \hline
    \end{tabular}}
\end{table}

\begin{figure}
    \includegraphics[width=\linewidth]{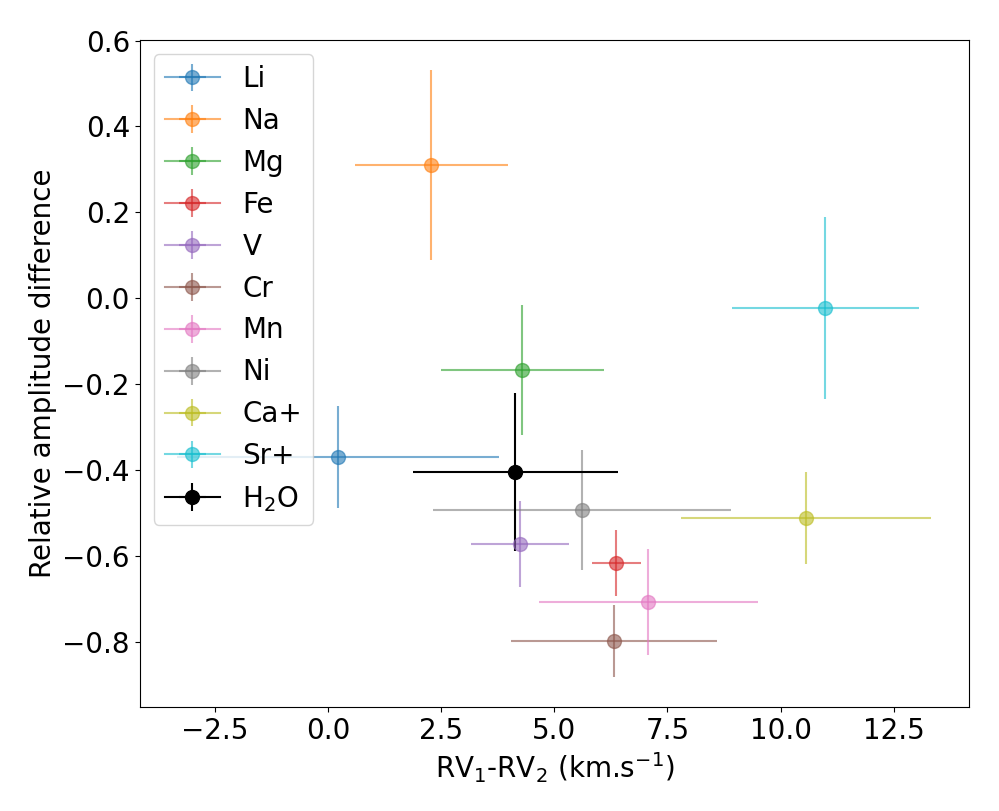}
    \caption{Scatter plot showing the relative amplitude difference as a function of radial velocity difference between the first and second half of the transit from \cite{Kesseli2022}, to which we have added our results for H$_2$O.}\label{fig:Kesseli_comp}
    
\end{figure}

To further investigate the asymmetry of H$_2$O between the limbs of WASP-76 b, we looked at each half of the transit separately, similarly to \cite{Gandhi2022} and \cite{Kesseli2022}. First, we separated the transit in two halves, splitting it in the middle, and performed the data reduction and cross-correlation procedures described in sections \ref{DataRed} and \ref{CCmaps_section} on each (see Appendix \ref{H2O_transithalves}). We were hence able to obtain K$_{\rm p}$ and V$_{\rm 0}$ values for each half of the transit (see Table \ref{tab:maxKpVsysfortransithalves_H2Omodels}), considering these to be first-order approximations of what would be obtained applying the same procedure as \cite{Gandhi2022} to the water signal. Our results for the first and second halves are consistent with those from \cite{Gandhi2022} for respectively the first and last quarters for the Fe signal. The Doppler shift obtained for the first half of the transit is consistent with what we previously obtained for the whole transit, with a blue-shift that could correspond to a significant contribution to the signal coming from both limbs. As for the second half, our measurement is consistent with a Doppler shift approximately twice as blue-shifted as for the first half, which could point to the signal being dominated by the trailing limb, where the combination of planetary rotation and day-to-night winds result in a greater blue-shift. To further our analysis, we also computed the mean cross correlation functions for each half from the 2D cross correlation function map, to perform a similar analysis as \cite{Kesseli2022}. We used the results of our Gaussian fits to each of these mean functions to compare our results for H$_2$O to those found by \cite{Kesseli2022}, in particular reproducing the comparison between relative amplitude differences as a function of radial velocity difference (see Figure \ref{fig:Kesseli_comp}). As before, our results are compatible with water having a similar asymmetry as Fe, with the signal from the trailing limb seeming stronger than that from the leading limb.

\subsubsection{CO}

For CO, we found a peak centred at a higher velocity than for H$_2$O, obtaining the maximum SNR for V$_{\rm 0}$ = -11.5 km.s$^{-1}$. This velocity is compatible with the results found for the trailing limb by \cite{Ehren2020} and \cite{Gandhi2022}, being within 1$\sigma$ of the velocity measured by \cite{Ehren2020}. This could imply that the overall signal for CO is dominated by that from the trailing limb of the planet, with the measured Doppler shift resulting from the combination of the planetary rotation and day-to-night winds on that side of WASP-76 b.

Though we already found indications as to the asymmetry of the CO signal, we attempted nevertheless to further our analysis by looking at each half of the transit separately, as we previously did for H$_2$O. However, we were unable to detect CO in either half separately. Additional data is required so as to have a sufficient SNR for each half to perform such a study for CO. We plan to revisit this analysis in a future work involving more transits observed by SPIRou of WASP-76 b.

\subsubsection{Comparing to GCMs}

\begin{figure}
    \includegraphics[width=\linewidth]{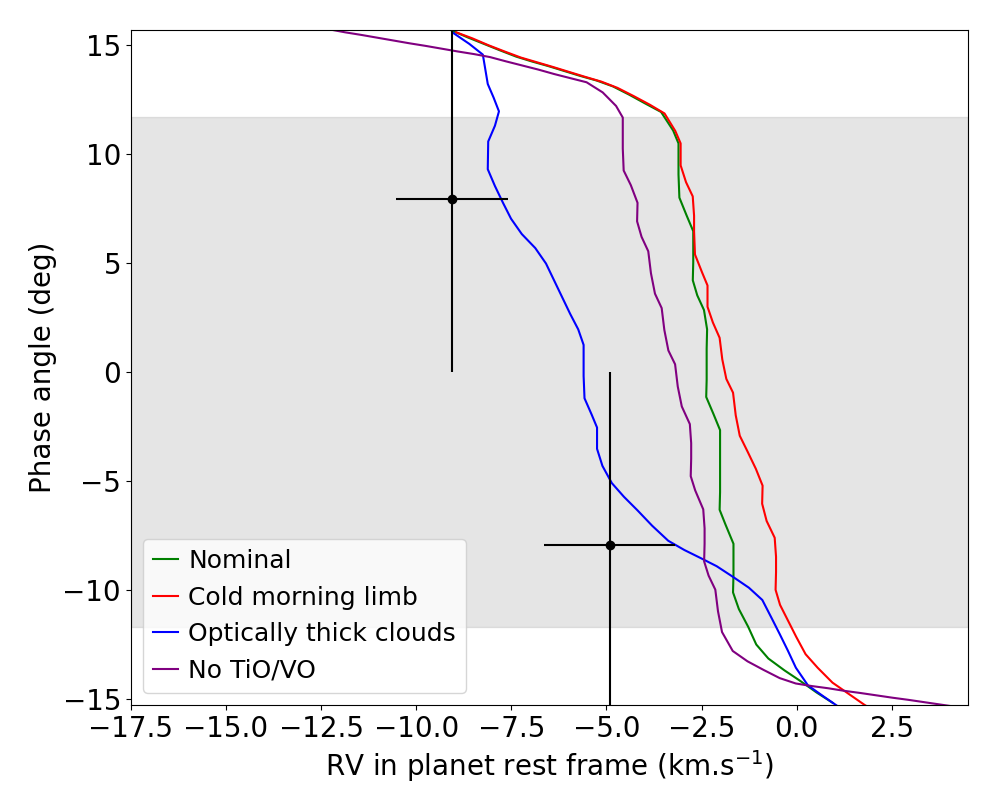}
    \caption{Comparison between the cross correlation profiles for different scenarios for H$_2$O obtained in \cite{Wardenier2023} and the radial velocities we obtained for H$_2$O from the mean cross correlation functions for each half of transit. The "nominal" profile (green) was obtained for a weak-drag model, the "cold morning limb" profile (red) for a model for which the temperature was reduced on the leading limb, the "optically thick clouds" profile (blue) for adding clouds à la \cite{Savel2022}, and the "no TiO/VO" profile (purple) for removing TiO and VO from the GCM calculations, considering them cold-trapped. The black dots correspond to the radial velocities obtained from Gaussian fits to the mean cross correlation functions for each half of transit, the horizontal black bars corresponding to the associated errors, and the vertical black bars indicate the associated phase range. The grey band indicates the transit duration from post-ingress to pre-egress.}\label{fig:GCM_watercomp}
    
\end{figure}

In \cite{Wardenier2023}, cross correlation maps were given for different species (including H$_2$O, CO and Fe) for different atmospheric models. We compared the resulting profiles to our results for H$_2$O and CO, as well as the results obtained by \cite{Ehren2020} and \cite{Kesseli2022} for Fe (as they use the same ESPRESSO dataset). For CO and Fe, the results were most compatible with the models where TiO and VO are cold-trapped and where there is a cold morning limb. As both of these species are expected to predominantly probe the dayside \citep{Wardenier2023} and both models have an asymmetry in temperature between limbs, these results point to a significant difference in temperature between the two limbs of WASP-76 b being behind the asymmetric signals of CO and Fe. As for H$_2$O, a species expected to predominantly probe the nightside \citep{Wardenier2023}, we found that our results were most consistent with the model involving optically thick clouds (see Figure \ref{fig:GCM_watercomp}). However, the strength of our H$_2$O signal as well as the results for clouds previously discussed in section \ref{metalclouds} could point to these optically thick clouds being lower in the atmosphere than for the model used by \cite{Wardenier2023} and/or these clouds being predominantly on the morning side of the planet. Note that due to the uncertainties on our results concerning H$_2$O and CO, we cannot yet confirm the indications given here from our results as to the scenario corresponding to the atmosphere of WASP-76 b. Nevertheless, our results do point to this scenario being more complex than the ones from \cite{Wardenier2023}, with it possibly being a combination of those different scenarios. 

\section{Conclusion}
\label{sec:concl}

We used the publicly available ATMOSPHERIX pipeline to analyse SPIRou acquired transmission spectra of one transit of the ultra-hot Jupiter WASP-76~b. Using models created with \texttt{\texttt{petitRADTRANS}}, we detected two major C- and O-bearing species, water and carbon monoxide, and measured their abundances local to the pressures probed by SPIRou for both cloudy and cloud-free models. Having included a grey cloud deck in our models, we found it to be favoured deep in the atmosphere, and measured log(H$_2$O)$_{\rm MMR}$ = -4.52 $\pm$ 0.77 and log(CO)$_{\rm MMR}$~= -3.09 $\pm$~1.05, leading to a C/O estimation of 0.94 $\pm$ 0.39 ($\sim$~1.7~$\pm$~0.7~$\times$~solar, with errors indicated corresponding to the 2$\sigma$ values). For a cloud-free model, though we measured slightly lower abundances for H$_2$O and CO than for with the inclusion of a grey cloud deck, we estimated a similar C/O ratio as for models with a grey cloud deck, finding C/O = 0.93 $\pm$ 0.34 ($\sim$ 1.7 $\pm$ 0.6 $\times$ solar). We also tried to detect HCN, C$_2$H$_2$ and OH, but we were only able to determine upper limits for their abundances. These were consistent with upper limits of our detection capabilities for these species determined using the data simulator of the ATMOSPHERIX analysis pipeline. These cloud-free models were also in agreement with a local C/O $\sim$ 0.9. To account for possible biases coming from the use of uniform distribution for the abundances of H$_2$O and CO as priors of our NS algorithm, we also used uniform distributions of C/O and [(C+O)/H] (a proxy for metallicity) as priors. However, we still found a preference for a C/O close to 1, obtaining a maximum likelihood for C/O = 0.90$_{-0.78}^{+0.07}$ ($\sim$ 1.64$_{-1.42}^{+0.13}$ $\times$ solar). Furthermore, with the abundance measurements of H$_2$O and CO and the upper limits found for HCN, C$_2$H$_2$ and OH, we investigate the validity of our C/O ratio for the pressures probed by SPIRou using \texttt{\texttt{ATMO}} models. Overall, we found that the abundance measurements of H$_2$O and CO and the upper limits determined for HCN, C$_2$H$_2$ and OH were consistent with a local C/O ratio between 1 and 1.8 $\times$ solar. This is thus consistent with our C/O ratio estimation for the pressures probed by SPIRou. However, the large uncertainties of our C/O estimation due to the large uncertainties of our CO abundance measurements preclude from reasonably placing constraints on the formation scenario of this planet. 

Considering the famous Fe asymmetric signal of this planet found by \cite{Ehren2020}, we also investigated the dynamics of the planet. While the initial detection signal for CO pointed to carbon monoxide having a similar asymmetry to the one for iron, further investigation into each half of the transit was required for H$_2$O. Nevertheless, we ultimately found an asymmetric signature for water that was also similar to the one for iron, though by comparing the results to GCM results, it became clear that these asymmetries could have different causes. Indeed, while the Fe and CO asymmetries indicated a temperature asymmetry, the explanation for the H$_2$O asymmetry pointed to the presence of clouds. However, due to the large uncertainties of our results, additional data is required to confirm exactly which scenario is behind the asymmetries that have been found for species detected in the atmosphere of WASP-76 b. Nevertheless, our results do indicate that it could be a combination of temperature asymmetry and presence of clouds. Future work with JWST observations to measure the full 3-D temperature structure and cloud cover of the planet could hence be key in understanding the asymmetric signatures of WASP-76 b.

We were able to study here the pressures probed by SPIRou at the full terminator of WASP-76~b, investigating the local C/O ratio and the dynamics of the corresponding pressure levels. By expanding this study to include data from other instruments, both ground- and space-based (such as JWST), acquired in transmission and emission, we could extend our local C/O estimation to one that is global to the planet. This could also lead to global estimations of other ratios such as Fe/O, used to constrain formation scenarios further than what is indicated by C/O.

%%%%%%%%%%%%%%%%%%%%%%%%%%%%%%%%%%%%%%%%%%%%%%%%%%%%%%%%%%%%%%%%%%%%%%%%%%%%%%%
%%%%%%%%%%%%%%%%%%%%%%%%%%%%%%%%%%%%%%%%%%%%%%%%%%%%%%%%%%%%%%%%%%%%%%%%%%%%%%%

\section*{Acknowledgements}
FD thanks the CNRS/INSU Programme National de Planétologie (PNP) and Programme National de Physique Stellaire (PNPS) for funding support. FD acknowledges funding from the Centre National d'Etudes Spatiales in the context of the French participation to the ARIEL mission. BK acknowledges funding from the European Research Council under the European Union’s Horizon 2020 research and innovation programme (grant agreement no. 865624, GPRV). E.M. acknowledges funding from FAPEMIG under project number APQ-02493-22 and research productivity grant number 309829/2022-4 awarded by the CNPq, Brazil. This work was supported by the Action Spécifique Numérique of CNRS/INSU. This work was granted access to the HPC resources of CALMIP supercomputing center under the allocation 2021-P21021.

%%%%%%%%%%%%%%%%%%%%%%%%%%%%%%%%%%%%%%%%%%%%%%%%%%
\section*{Data Availability}

\bibliographystyle{aa} % style aa.bst
\bibliography{main}

%\begin{thebibliography}{}
%\end{thebibliography}

%%%%%%%%%%%%%%%%% APPENDICES %%%%%%%%%%%%%%%%%%%%%

\appendix

\section{Table of chemical species in WASP-76~b}
In Table \ref{tab:chemlist_table}, we show a list of chemical species that have been included in previous studies of the atmosphere of WASP-76~b. For each species, we indicate the instrument used to study its presence in the atmosphere of WASP-76~b, give the reference to the paper presenting the study, and if it was reported as detected (highlighted green), tentatively detected (highlighted orange) or non-detected (highlighted red).

\begin{sidewaystable*}
%\begin{landscape}
%    \begin{table}
    %\vspace{4.2cm}
    \centering
    \begin{tabular}{|c|c|p{0.025\textwidth}p{0.025\textwidth}|c||p{0.06\textwidth}p{0.06\textwidth}|c|c|c|c|c|c|c|}\cline{3-14}
    %\begin{tabular}{|p{0.1\textwidth}|c|c|c||cc|c|c|c|c|c|c|c|c|}\cline{3-14}
         \multicolumn{1}{c}{}& \multicolumn{1}{c|}{}& \multicolumn{3}{c||}{Space-based observations} & \multicolumn{9}{c|}{Ground-based observations}\\ \cline{3-14}
         \multicolumn{1}{c}{} & \multicolumn{1}{c|}{} & \multicolumn{2}{c|}{HST} & Spitzer & \multicolumn{2}{c|}{ESPRESSO} & GRACES & HARPS & Subaru/HDS & MAROON-X & CARMENES & CRIRES+ & SPIRou \\\hhline{--|==|=||==|=|=|=|=|=|=|=|}
         \multirow{16}{*}{\rotatebox[origin=c]{90}{Atoms}} & H &  &  & & \multicolumn{2}{c|}{8, 14, 15\cellcolor{orange!80}} & & 4\cellcolor{candyapplered!80} &    & 20\cellcolor{brightgreen!60} & 12\cellcolor{candyapplered!80} &  &\\\cline{2-14}%\hline
         & He & & & & &  &  &  &  &  & 12\cellcolor{orange!80} & & \\\cline{2-14}
         & Li & & &  & \multicolumn{2}{c|}{8, 14, 15\cellcolor{brightgreen!60}} & 18\cellcolor{orange!80} &  &  & 20\cellcolor{brightgreen!60} & 12\cellcolor{candyapplered!80} & & \\\cline{2-14}
         & O & & & &  \multicolumn{2}{c|}{14\cellcolor{candyapplered!80}}  & 18\cellcolor{candyapplered!80} &  &  & 20\cellcolor{orange!80} &  & & \\\cline{2-14}
         & Na & \multicolumn{2}{c|}{5\cellcolor{brightgreen!60}}  & & 8, 14, 15 \centering \cellcolor{brightgreen!60} & 22 \centering \cellcolor{candyapplered!80} & 18\cellcolor{brightgreen!60} & 3, 4\cellcolor{brightgreen!60} & 16\cellcolor{brightgreen!60} & 20\cellcolor{brightgreen!60} & 12\cellcolor{candyapplered!80} & & \\\cline{2-14}
         & Mg & & &  & \multicolumn{2}{c|}{8, 14, 15, 22\cellcolor{brightgreen!60}} & 18\cellcolor{candyapplered!80} &  &  & 20\cellcolor{brightgreen!60} &  & & \\\cline{2-14}
         & Al & & & & \multicolumn{2}{c|}{14\cellcolor{candyapplered!80}}   & 18\cellcolor{candyapplered!80} &  &  & 20\cellcolor{candyapplered!80} &  & & \\\cline{2-14}
         & K & & &  & 8\centering \cellcolor{brightgreen!60}& 14 \centering\cellcolor{orange!80}& 18\cellcolor{orange!80} &  &  & 20\cellcolor{brightgreen!60} & 12\cellcolor{candyapplered!80} & &\\\cline{2-14}
         & Ca & \multicolumn{2}{c|}{9\cellcolor{orange!80}} & & 22 \centering\cellcolor{brightgreen!60} & 14, 15 \centering\cellcolor{candyapplered!80} &  18\cellcolor{candyapplered!80} &  &  & 20\cellcolor{brightgreen!60} &  & &\\\cline{2-14}
         & Sc & & & & \multicolumn{2}{c|}{14\cellcolor{candyapplered!80}}   & 18\cellcolor{candyapplered!80} &  &  & 20\cellcolor{candyapplered!80} &  & & \\\cline{2-14}
         & Ti & \multicolumn{2}{c|}{9\cellcolor{orange!80}} & & \multicolumn{2}{c|}{8, 14, 22 \cellcolor{candyapplered!80}}   & 18\cellcolor{candyapplered!80} &  &  & 20\cellcolor{candyapplered!80} &  & & \\\cline{2-14}
         & V & & &  & \multicolumn{2}{c|}{14, 15, 22\cellcolor{brightgreen!60}} & 18\cellcolor{orange!80} &  &  & 20\cellcolor{brightgreen!60} &  & & \\\cline{2-14}
         & Cr & & & &14, 15, 22 \centering\cellcolor{brightgreen!60} & 8\centering \cellcolor{candyapplered!80} & 18\cellcolor{orange!80} &  &  & 20\cellcolor{brightgreen!60} &  & & \\\cline{2-14}
         & Mn & & &  & \multicolumn{2}{c|}{8, 14, 15, 22\cellcolor{brightgreen!60}} & 18\cellcolor{candyapplered!80} &  &  & 20\cellcolor{brightgreen!60} &  & & \\\cline{2-14}
         & Fe & \multicolumn{2}{c|}{9\cellcolor{orange!80}} &  & \multicolumn{2}{c|}{7, 8, 14, 15, 17, 22\centering \cellcolor{brightgreen!60}} & 18\cellcolor{brightgreen!60} & 10\cellcolor{brightgreen!60}  \cellcolor{brightgreen!60} &  & 20\cellcolor{brightgreen!60} &  & & \\\cline{2-14}
         & Co & & &  & 14 \centering\cellcolor{orange!80}& 15\centering\cellcolor{candyapplered!80} & 18\cellcolor{candyapplered!80} &  &  &  &  & & \\\cline{2-14}
         & Ni & \multicolumn{2}{c|}{9\cellcolor{orange!80}} & & 14, 22\centering\cellcolor{brightgreen!60} & 8, 15\centering \cellcolor{candyapplered!80} & 18\cellcolor{candyapplered!80} &  &  & 20\cellcolor{brightgreen!60} &  & & \\\cline{2-14}
         & Zr&  & & & \multicolumn{2}{c|}{14\cellcolor{candyapplered!80}}  & 18\cellcolor{candyapplered!80} &  &  &  &  &  &\\\hline
         \multirow{6}{*}{\rotatebox[origin=c]{90}{Ions}} & Ca+ & \multicolumn{2}{c|}{9\cellcolor{orange!80}} & & \multicolumn{2}{c|}{8, 14, 15 \cellcolor{brightgreen!60}} & 11, 18\cellcolor{brightgreen!60} &  &  & 20\cellcolor{brightgreen!60} & 12\cellcolor{brightgreen!60} & & \\\cline{2-14}
         & Sc+& & & & \multicolumn{2}{c|}{14\cellcolor{candyapplered!80}}  & 18\cellcolor{candyapplered!80} &  &  &  &  & & \\\cline{2-14}
         & Ti+&  & & & \multicolumn{2}{c|}{14\cellcolor{candyapplered!80}}  & 18\cellcolor{candyapplered!80} &  &  & 20\cellcolor{candyapplered!80} &  &  &\\\cline{2-14}
         & V+& & & &  \multicolumn{2}{c|}{14\cellcolor{candyapplered!80}}  &  &  &  & 20\cellcolor{candyapplered!80} &  & & \\\cline{2-14}
         & Fe+& \multicolumn{2}{c|}{9\cellcolor{orange!80}} &  & \multicolumn{2}{c|}{15, 18\cellcolor{candyapplered!80}} &  &  &  & 20\cellcolor{orange!80} &  & & \\\cline{2-14}
         & Sr+& &  &  & 14\centering\cellcolor{brightgreen!60}& 15\centering\cellcolor{candyapplered!80}& 18\cellcolor{candyapplered!80} &  &  &  &  & & \\\cline{2-14}
         & Ba+ &&  &  & \multicolumn{2}{c|}{15\cellcolor{brightgreen!60}} & 18\cellcolor{candyapplered!80} &  &  & 20\cellcolor{brightgreen!60} &  &  &\\\hline
         \multirow{12}{*}{\rotatebox[origin=c]{90}{Molecules}} & CH$_4$  & \multicolumn{2}{c|}{1, 6\cellcolor{candyapplered!80}} & &  &  &  &  &  &  & 19\cellcolor{candyapplered!80} &  &\\\cline{2-14}
         & OH && & &  &  &  &  &  &  & 13, 24\cellcolor{brightgreen!60} &  &25\cellcolor{candyapplered!80}\\\cline{2-14}
         & NH$_3$ &\multicolumn{2}{c|}{ 1, 2\cellcolor{candyapplered!80}} & &  &  &  &  &  &  & 19\cellcolor{orange!80}  & &\\\cline{2-14}
         & H$_2$O & \multicolumn{2}{c|}{1, 2, 6, 23\cellcolor{brightgreen!60}}  & & & & 18\cellcolor{candyapplered!80} &  &  &  & 19\cellcolor{brightgreen!60}  & 21\cellcolor{orange!80} & 25\cellcolor{brightgreen!60} \\\cline{2-14}
         & C$_2$H$_2$ && & &  &  &  &  &  &  &  & & 25\cellcolor{candyapplered!80}\\\cline{2-14}
         & HCN & \multicolumn{2}{c|}{2\cellcolor{candyapplered!80}}& &  &  &  &  &  &  & 19\cellcolor{brightgreen!60}  & &25\cellcolor{candyapplered!80}\\\cline{2-14}
         & CO & \multicolumn{2}{c|}{1, 6\cellcolor{candyapplered!80}}&  9\cellcolor{brightgreen!60} & & & 18\cellcolor{candyapplered!80} &  &  &  & 19\cellcolor{candyapplered!80} & 21\cellcolor{brightgreen!60} & 25\cellcolor{brightgreen!60}\\\cline{2-14}
         & AlO && & &  &  &  &  &  &   20\cellcolor{candyapplered!80} & & & \\\cline{2-14}
         & CO$_2$ & \multicolumn{2}{c|}{1, 6\cellcolor{candyapplered!80}} & &  &  &  &  &  &  & 19\cellcolor{candyapplered!80} & & \\\cline{2-14}
         & SiO & \multicolumn{2}{c|}{9\cellcolor{orange!80}}  & &  &  &  &  &  &  &  & & \\\cline{2-14}
         & CrH& & &   &  &  &  &  &  & 20\cellcolor{candyapplered!80} & & &\\\cline{2-14}
         & FeH& \multicolumn{2}{c|}{6\cellcolor{orange!80}} & &  &  &  &  &  &  &  &  &\\\cline{2-14}
         & TiO &  1, 6 \centering\cellcolor{brightgreen!60}\centering & 6\centering\cellcolor{orange!80} & & \multicolumn{2}{c|}{8, 22 \cellcolor{candyapplered!80} }  &  &  &  & 20\cellcolor{candyapplered!80} &  & & \\\cline{2-14}
         & VO &  1\centering\cellcolor{brightgreen!60}\centering& 6\centering\cellcolor{orange!80} & & \multicolumn{2}{c|}{8 \cellcolor{candyapplered!80}}  &  &  &  & 20\cellcolor{brightgreen!60} &  & & \\\cline{2-14}
         & ZrO& & & & \multicolumn{2}{c|}{8 \cellcolor{candyapplered!80}}  &  &  &  &  &  & & \\\hline
    \end{tabular}
    \caption{List of chemical species reported as detected (green), tentatively detected (orange), or not detected (red) in previous publications and this work. For each species has been indicated the instrument used and the corresponding reference for the study of its presence in the atmosphere. Empty cells indicate that the species has not been explicitly included in a published study with the corresponding instrument.
    }
    \vspace{-0.6cm}
    \tablebib{(1)\citet{Tsiaras2018};(2)\citet{Fisher2018};(3)\citet{Seidel2019};(4)\citet{Zak2019};(5)\citet{vonEssen2020};(6)\citet{Edwards2020};(7)\citet{Ehren2020};(8)\citet{Tabernero2021};(9)\citet{Fu2021};(10)\citet{Kesseli2021};(11)\citet{Deibert2021};(12)\citet{CasasayasBarris2021};(13)\citet{Landman2021};(14)\citet{Kesseli2022}\footnote[1]{Due to the large number of species reported as non-detected, only the non-detections of species included in other studies are included in this list.\label{refnote}};(15)\citet{AzevedoSilva2022};(16)\citet{Kawauchi2022};(17)\citet{Gandhi2022};(18)\citet{Deibert2023}\footref{refnote};(19)\citet{SanchezLopez2022};(20)\citet{Pelletier2023};(21)\citet{Yan2023};(22)\citet{Gandhi2023};(23)\citet{Edwards2023};(24)\citet{Cheverall2023};(25)This work.}
    
    \label{tab:chemlist_table}
%    \end{table}
%\end{landscape}
\end{sidewaystable*}

%\definecolor{darkpastelgreen}{rgb}{0.01,0.75,0.24}
%\definecolor{lightgreen}{rgb}{0.56, 0.93, 0.56}
%\definecolor{brightgreen}{rgb}{0.4, 1.0, 0.0}
%\definecolor{carrotorange}{rgb}{0.93, 0.57, 0.13}
%\definecolor{darkorange}{rgb}{1.0, 0.55, 0.0}
%\definecolor{bostonuniversityred}{rgb}{0.8, 0.0, 0.0}
%\definecolor{cadmiumred}{rgb}{0.89, 0.0, 0.13}

\section{Rest frame absorption signals}
\label{restframe_abssignals}

Here are shown the planetary rest frame absorption trails for H$_2$O and CO.

\begin{figure}[ht]
    \centering
    \hspace{-0.4cm}\includegraphics[width=0.97\linewidth]{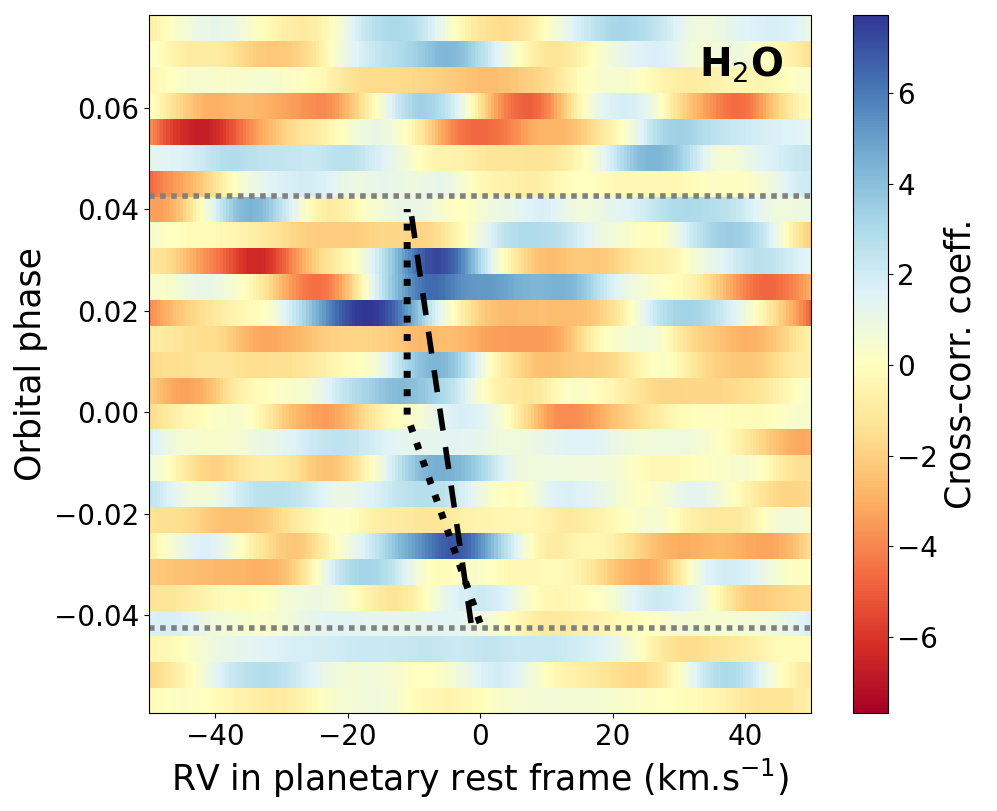}

    \includegraphics[width=\linewidth]{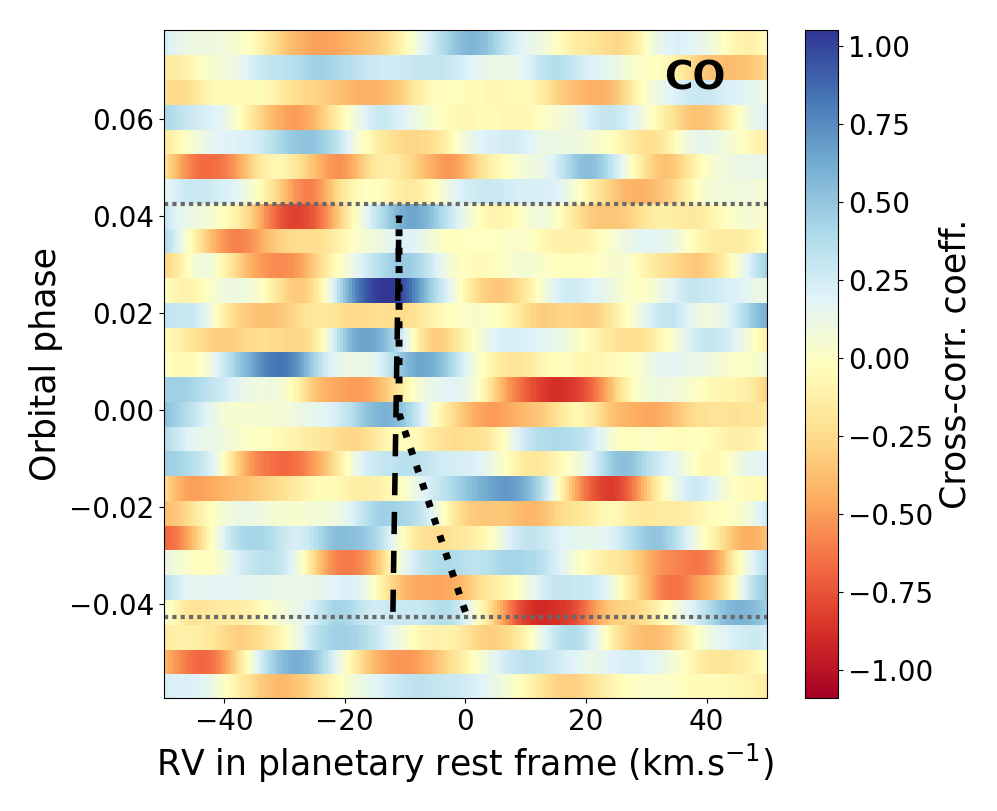}

    \caption{Cross-correlation trails for H$_2$O (top) and CO (bottom). The dashed lines correspond to the absorption trails indicated for each species by the K$_{\rm p}$ and V$_{\rm 0}$ values reported in section \ref{H2O&COdetectest}. The black dotted lines correspond to the "kinked" absorption trail reported for Fe. The grey dotted horizontal lines correspond to the beginning and end of the transit.}
    \label{fig:COtrail}
\end{figure}

\section{Changing priors} \label{changing priors}

As \texttt{\texttt{petitRADTRANS}} requires the abundance profile of a species when creating model spectra, we had to derive the expressions to calculate the MMRs of H$_2$O and CO from the expressions of the C/O and [(C+O)/H] (used as a proxy for metallicity). Considering only H$_2$, H$_2$O and CO present in the atmosphere (equivalent to considering all other species negligible in the calculation of these ratios), we derived from these two ratios the following:
\begin{equation}
    X_{i} = \frac{\frac{C+O}{H}}{\Gamma_{i}}
\end{equation}
with X$_i$ being the MMR of each species \textit{i} included in the list of line absorbers, corresponding here to H$_2$O and CO, which have the following expressions for $\Gamma_{i}$:

\begin{multline} \label{eq:H2O_priorcheck}
    \Gamma_{H_2O} = \frac{\mu_{H_2}}{2\mu_{H_2O}(1-\frac{H}{He})} + \frac{C+O}{H} \times \Bigg( 1- \frac{\mu_{H_2}}{\mu_{H_2O}(1-\frac{H}{He})} \\
    + \frac{C}{O} \times \frac{\mu_{CO}}{\mu_{H_2O}} \times \frac{\frac{\mu_{H_2}}{\mu_{CO} (1-\frac{H}{He})}+\frac{C+O}{H}}{1-\frac{C}{O}} \Bigg)        
\end{multline}

\begin{multline} \label{eq:CO_priorcheck}
    \Gamma_{CO} = \frac{\mu_{H_2}}{\mu_{CO}(1-\frac{H}{He})} + \frac{C+O}{H} +  \frac{1-\frac{C}{O}}{\frac{C}{O} \times \frac{\mu_{CO}}{\mu_{H_2O}}} \\
    \times \Bigg[ \frac{\mu_{H_2}}{2\mu_{H_2O}(1-\frac{H}{He})} + \frac{C+O}{H} \times \Bigg(1-\frac{\mu_{H_2}}{\mu_{H_2O}(1-\frac{H}{He})}\Bigg)\Bigg]
\end{multline}
with $\mu_i$ being the molecular mass of the species i (H$_2$O, CO or H$_2$), and using the solar H/He ratio ($\simeq$0.275).

\begin{figure}
    \includegraphics[width=\linewidth]{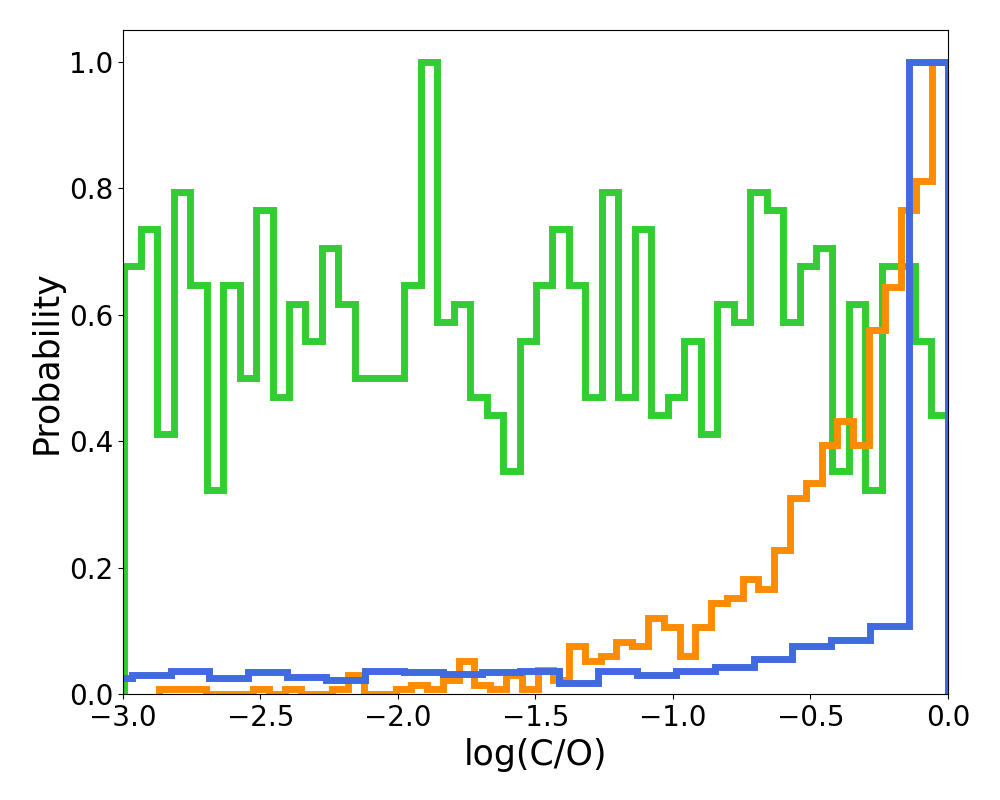}

    \includegraphics[width=\linewidth]{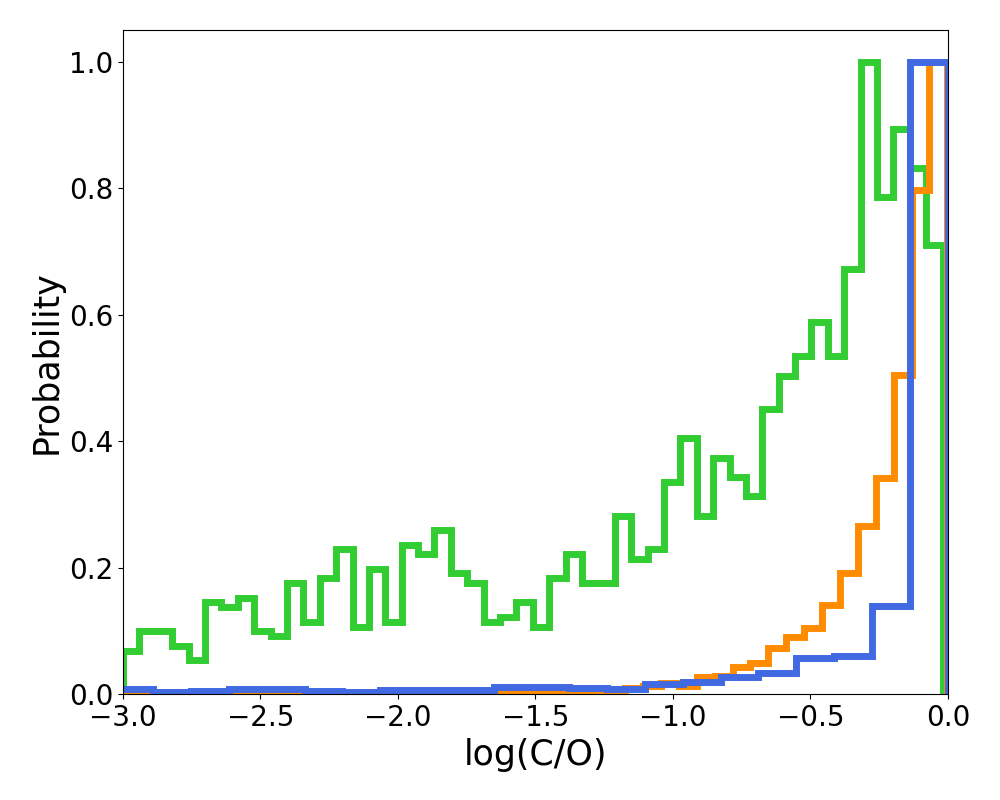}

    \caption{Distribution of log(C/O), with the top panel representing the priors and the bottom panel the posteriors. The blue lines correspond to when using uniform priors for the distributions of the MMRs of H$_2$O and CO. The orange lines correspond to when using a uniform linear distribution of the C/O ratio. The green lines correspond to using a uniform distribution in log-space of the C/O ratio.}
    \label{fig:compCOratio_prior_post}
\end{figure}

The NS results can be seen in Figure \ref{fig:CpOratCOratTeqPcloud_correctvrot}. We calculated the abundance probability distributions for H$_2$O and CO from those of C/O and [(C+O)/H], finding a maximum probability for log(H$_2$O)$_{\rm MMR}$ = -4.27 $\pm$ 0.54 and log(CO)$_{\rm MMR}$ = -3.48 $\pm$ 0.87. These values are both within 1$\sigma$ of those retrieved in section \ref{H2O&COdetectest}.

We were also able to use a uniform prior of log(C/O) by replacing C/O with log(C/O) in equations \ref{eq:H2O_priorcheck} and \ref{eq:CO_priorcheck}. In Figure \ref{fig:compCOratio_prior_post}, we compare the log(C/O) distributions of both the priors and posteriors of each of our 3 cases (using as priors uniform distributions of H$_2$O and CO, of C/O and [(C+O)/H], and of log(C/O) and [(C+O)/H]). We can see that only when using log(C/O) as prior do we not favour as a posterior log(C/O) $\sim$ 0. However, by using a uniform distribution of log(C/O) as a prior we also introduce a strong bias towards sub-solar values of C/O, and even extremely sub-solar values of C/O. Despite this strong bias, we still find in this case that a solar value for C/O is most likely. Taking into account the strong bias introduced by the chosen prior distribution, this seems to indicate a high likelihood for the C/O to be super-solar, in agreement with what was found in our other two cases of study.

\section{Water detections in transit halves} \label{H2O_transithalves}

\begin{figure}
    \includegraphics[width=\linewidth]{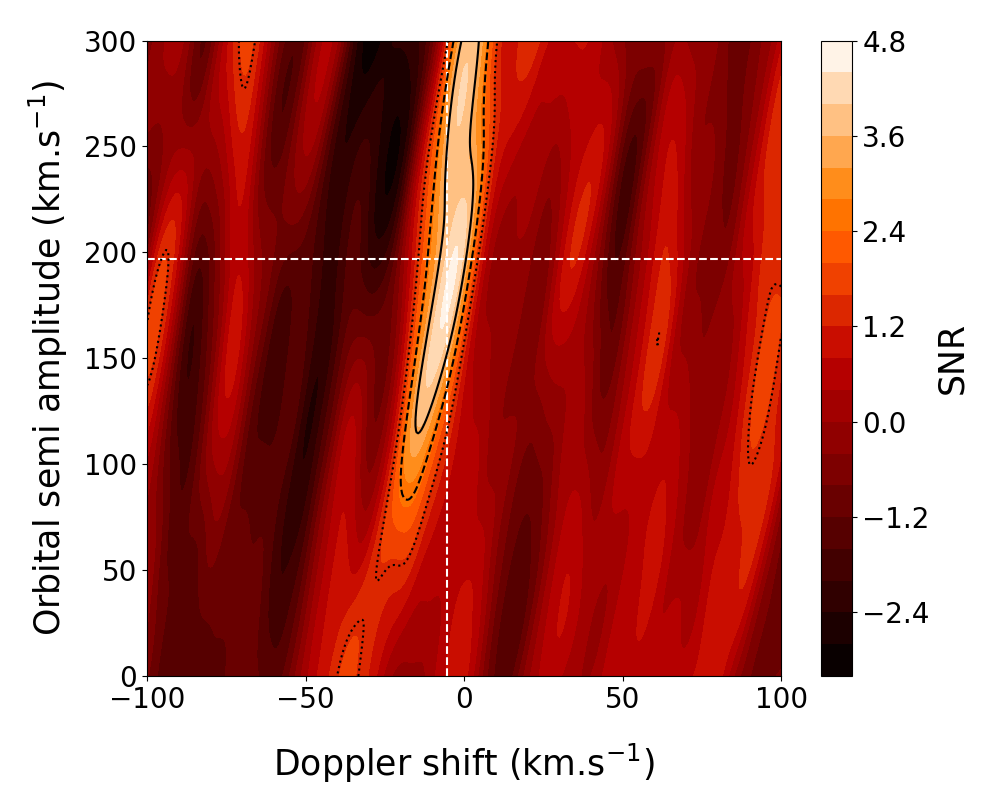}
    \caption{Map resulting from cross-correlation between the reduced data of WASP-76~b from the first half of the transit and spectra from a model containing H$_2$O opacity lines (log(H$_2$O)$_{\rm MMR}$ = -5.0), with T = 1500 K. The SNR varies from -3.16 to 4.65, with the maximum SNR of 4.65 obtained for K$_{\rm p}$ = 184 km.s$^{-1}$ and V$_{\rm 0}$ = -4.5 km.s$^{-1}$.} 
    \label{fig:waterhalf1}
\end{figure}

\begin{figure}
    \includegraphics[width=\linewidth]{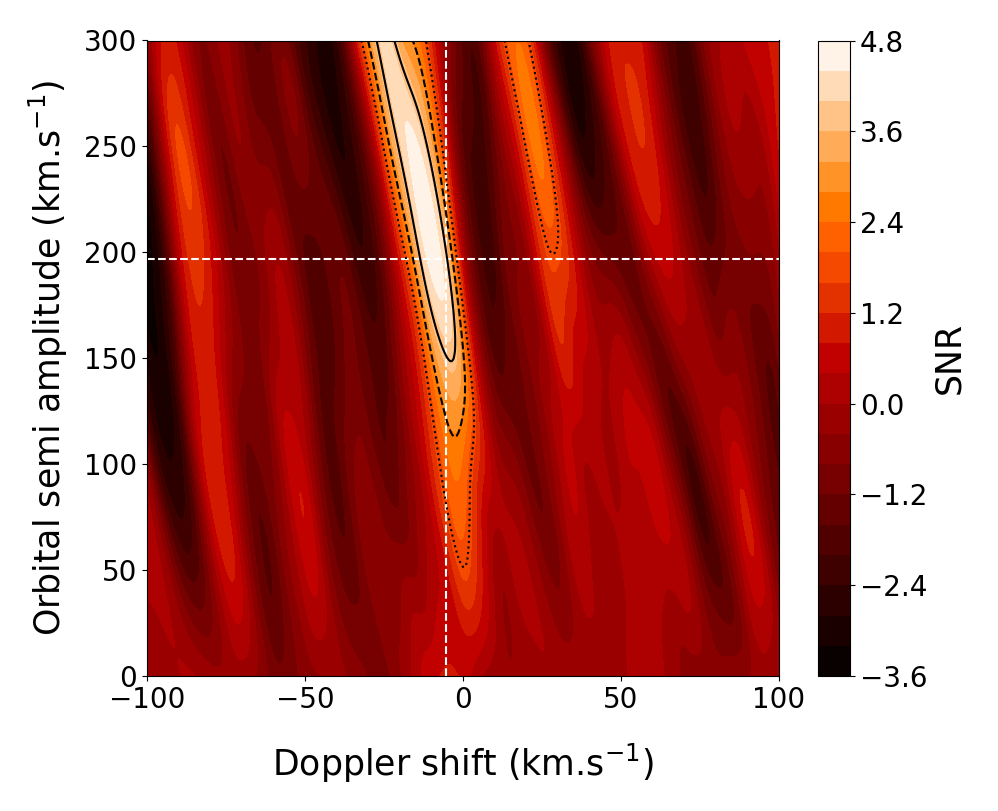}
    \caption{Map resulting from cross-correlation between the reduced data of WASP-76~b from the second half of the transit and spectra from a model containing H$_2$O opacity lines (log(H$_2$O)$_{\rm MMR}$ = -5.0), with T~=~1500~K. The SNR varies from -3.39 to 4.75, with the maximum SNR of 4.75 obtained for K$_{\rm p}$ = 216 km.s$^{-1}$ and V$_{\rm 0}$ = -11.5 km.s$^{-1}$.}
    \label{fig:waterhalf2}
\end{figure}

Here are shown the maps resulting from the cross-correlation between each half of the transit of WASP-76~b and cloud-free models with T = 1500 K in which we included only H$_2$O. Figure \ref{fig:waterhalf1} represents the resulting map for the first half of the transit. Figure \ref{fig:waterhalf2} represents the resulting map for the second half of the transit. 

\section{Retrieval results for H$_2$O and CO abundance estimation}
\label{apen:retrievalwithclouds}

Here are shown the retrieval results associated to section \ref{H2O&COdetectest}.

\begin{figure*}
    \includegraphics[width=\textwidth]{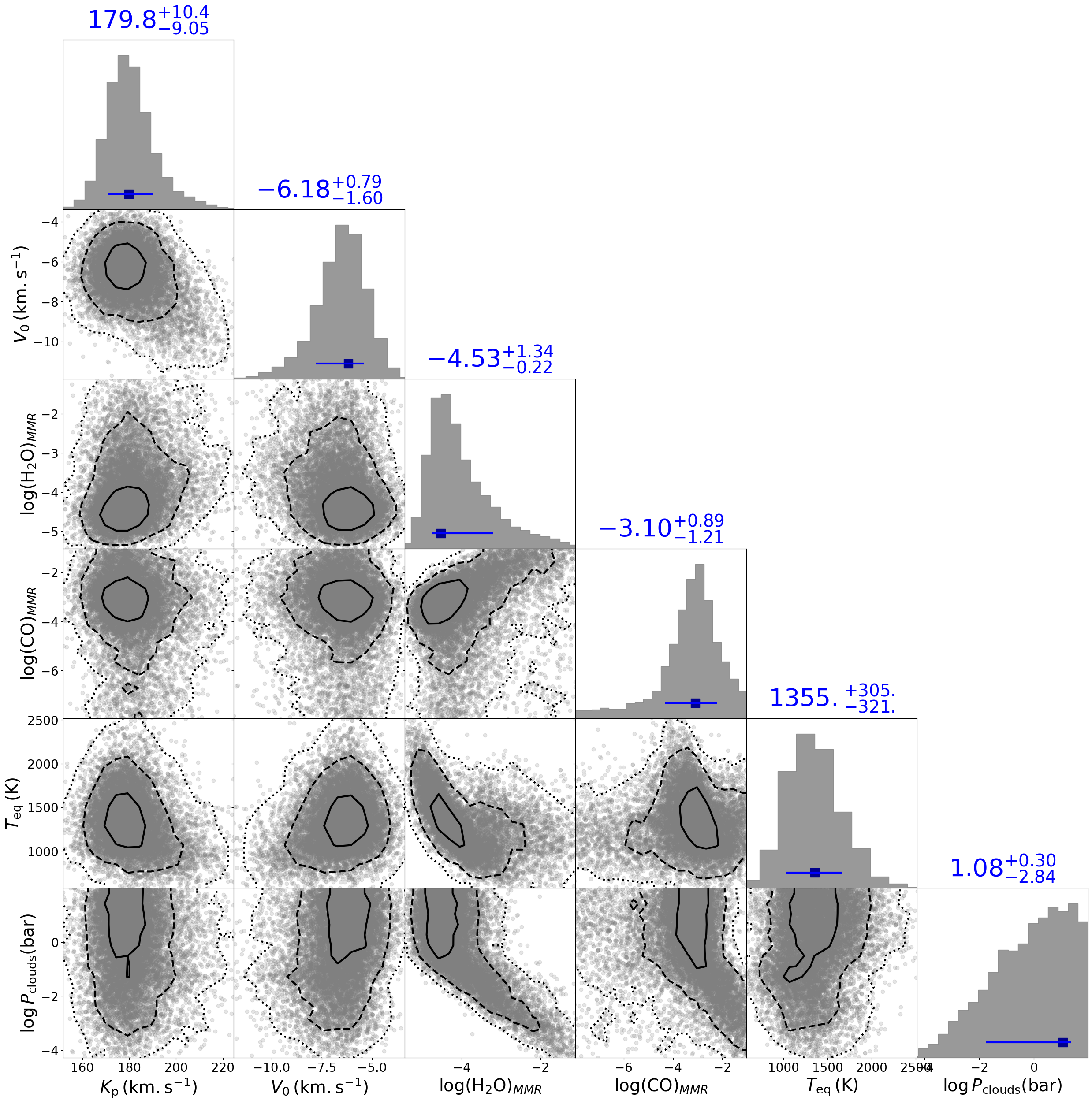}
    \caption{Corner plot showing the NS results of our retrieval using the WASP-76~b data. The black lines (full, dashed and dotted) respectively represent the 1, 2 and 3 sigma contours. The maximum probability and 1$\sigma$ values found are represented for each histogram on the diagonal by the blue dot and bar.}
    \label{fig:retrievalwithclouds}
\end{figure*}

\section{Retrieval results for v$_{\rm eq}$ estimation}
\label{apen:retrievalwithvrot}

Here are shown the retrieval results associated to section \ref{dynamics}.

\begin{figure*}
    \includegraphics[width=\textwidth]{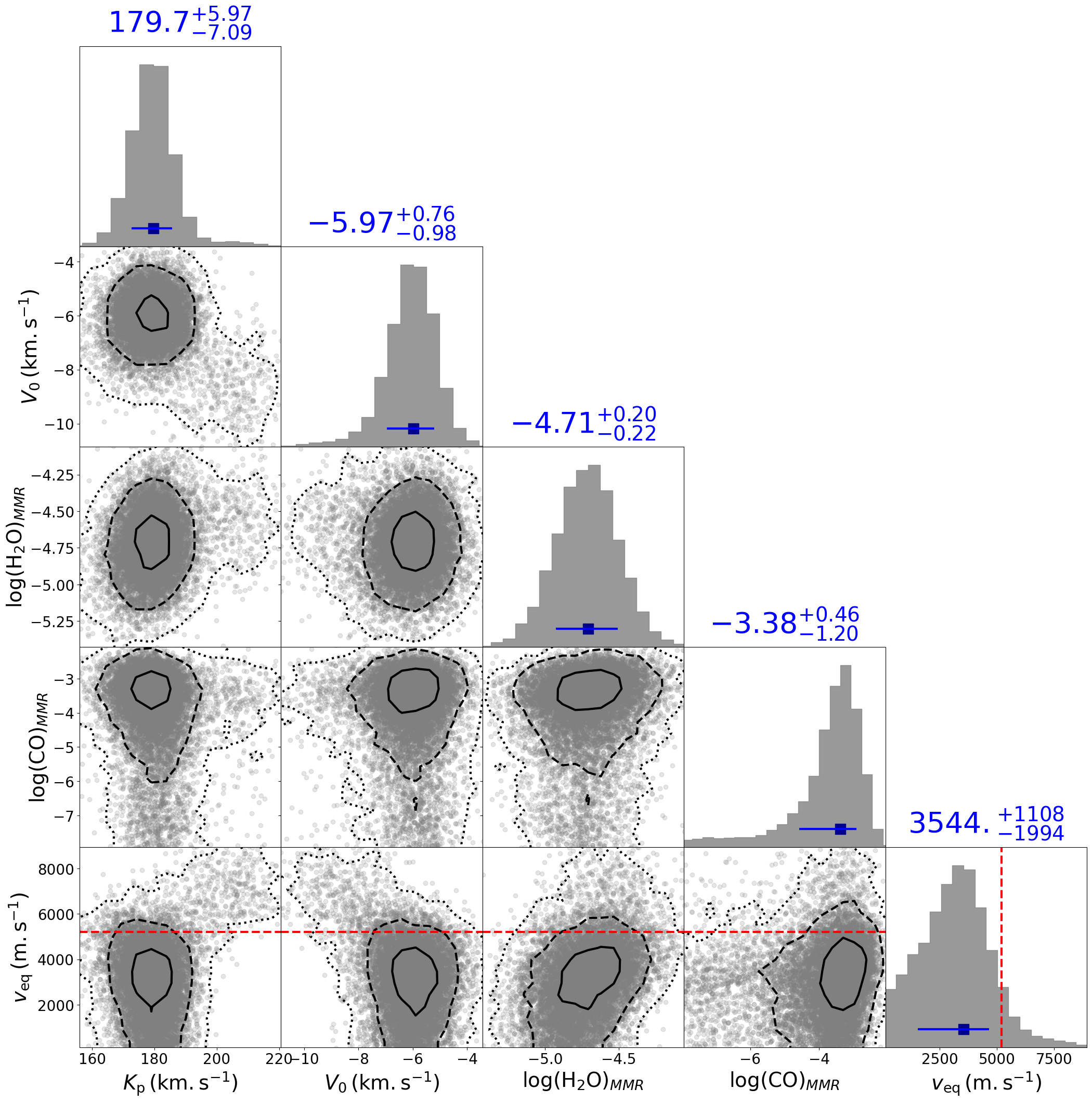}
    \caption{Posterior distributions obtained from the NS algorithm using cloud-free models with T = 1500K. Represented by the red dashed line is the expected rotation velocity for WASP-76~b (see section \ref{dynamics}).}
    \label{fig:retrievalH2O&CO}
\end{figure*}

\section{Retrieval results for C/O ratio and metallicity}
\label{apen:retrievalwithCOrat&metal}

Here are shown the retrieval results associated to section \ref{sssec:C/O robustness}.

\begin{figure*}
    \includegraphics[width=\textwidth]{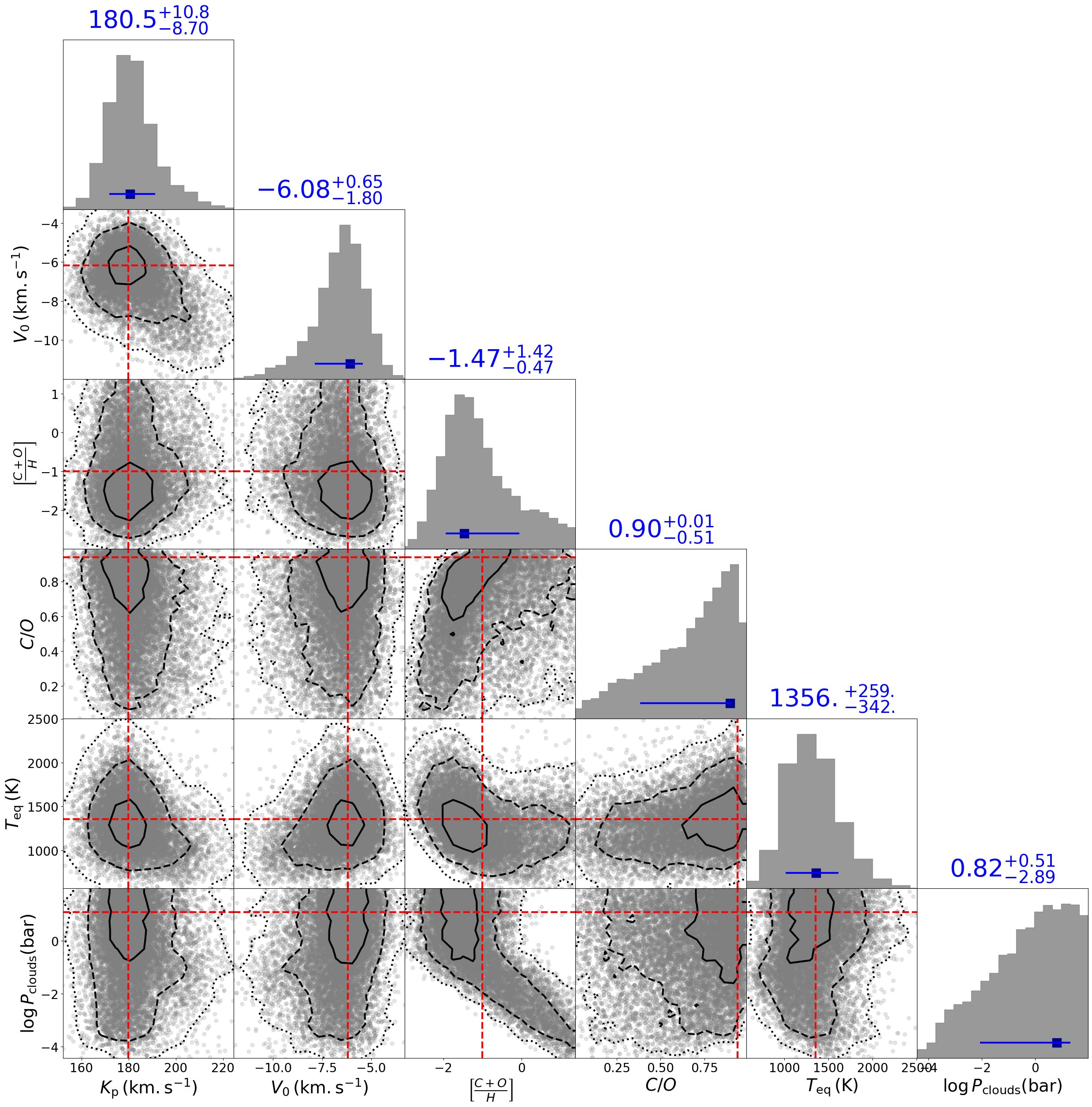}
    \caption{NS posterior distribution for using C/O and [(C+O)/H] as priors. The red dashed lines represent the values found in Figure \ref{fig:retrievalwithclouds}}
    \label{fig:CpOratCOratTeqPcloud_correctvrot}
\end{figure*}

\section{Retrievals with HCN and C$_2$H$_2$, and OH}
\label{C}

\begin{figure*}
    \includegraphics[width=\textwidth]{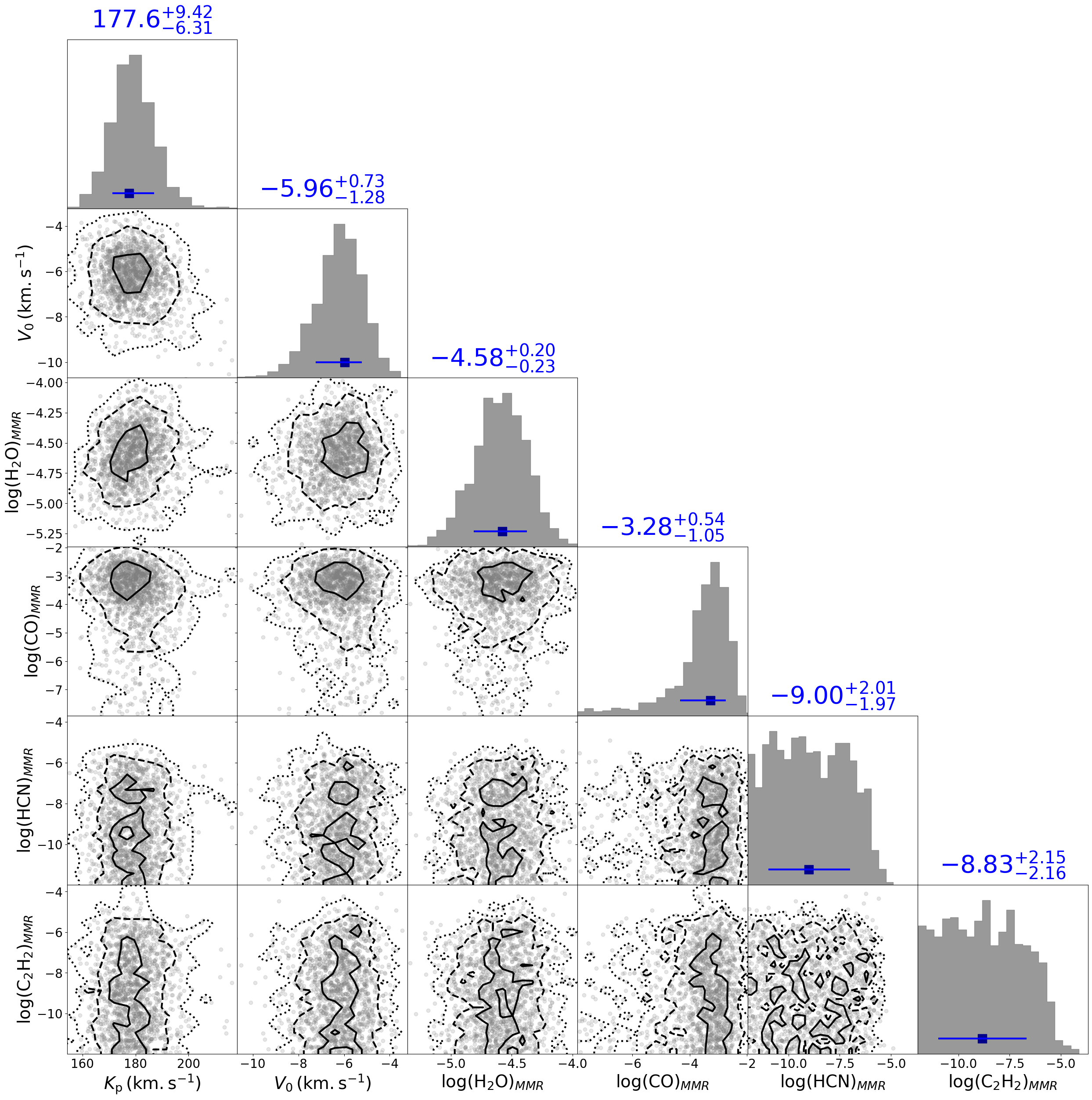}
    \caption{NS obtained posterior distributions for an atmosphere with T = 1500K and v$_{\rm rot}$=5210 m.s$^{-1}$, and for which we looked for HCN and C$_2$H$_2$. }
    \label{fig:retrievalwithHCN&C2H2}
\end{figure*}

\begin{figure*}
    \includegraphics[width=\textwidth]{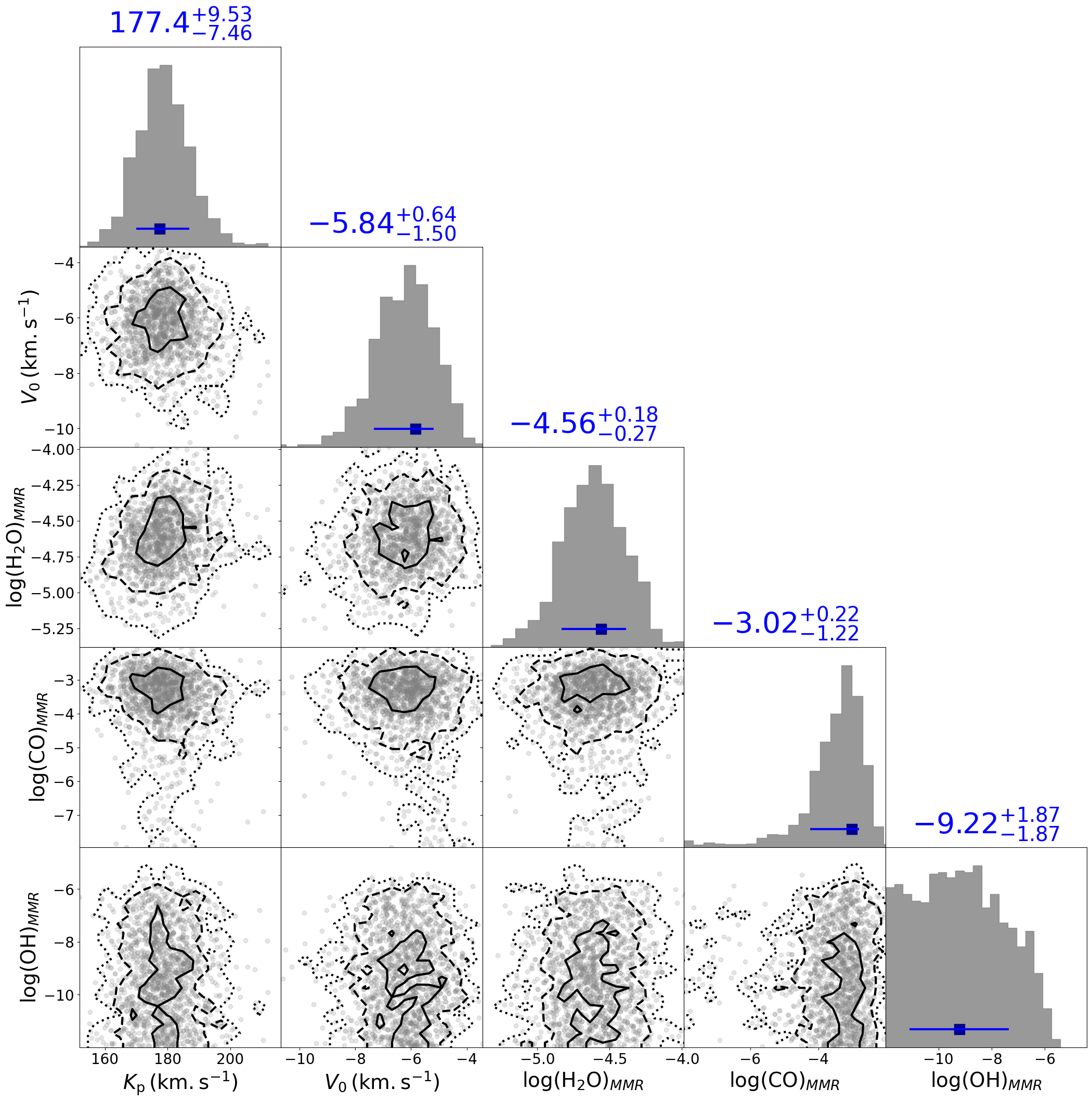}
    \caption{NS obtained posterior distributions for an atmosphere with T = 1500K and v$_{\rm rot}$ = 5210 m.s$^{-1}$, and for which we looked for OH.}
    \label{fig:retrievalwithOH}
\end{figure*}

Here are shown the retrieval results for cloud-free models with T=1500K and in which we included H$_2$O, CO. For Figure \ref{fig:retrievalwithHCN&C2H2}, we also included HCN and C$_2$H$_2$, and for Figure \ref{fig:retrievalwithOH}, we also included OH.

\end{document}